%% file: dalitz-spires.tex
\newcount\MiniFinal \MiniFinal = 0
\newcount\Miniwww \Miniwww = 3
\newwrite\pageindexwrite
\input mtexsis
\input minimacros
\let\Input=\input

\beginmini
\referencelist

\reference{cleoa}
H.~Muramatsu {\it et al.}
[CLEO Collaboration],
Phys.\ Rev.\ Lett.\  {\bf 89}, 251802 (2002)
[Erratum-ibid.\  {\bf 90}, 059901 (2003)]
\endreference

\reference{cleob}
G.~Brandenburg {\it et al.} 
[CLEO Collaboration], 
Phys.\ Rev.\ Lett. {\bf 87}, 071802 (2001)
\endreference

\reference{cleopipipi0}
V.~V.~Frolov {\it et al.} [CLEO Collaboration],
arXiv:hep-ex/0306048. Contributed to the International Europhysics Conference on High Energy Physics (EPS 2003)
\endreference

\reference{e687pipipi}
P.~L.~Frabetti {\it et al.} [E687 Collaboration],
Phys.\ Lett.\ B {\bf 407}, 79 (1997)
\endreference

\reference{e791dspipipi}
E.~M.~Aitala {\it et al.} [E791 Collaboration],
Phys.\ Rev.\ Lett. {\bf 86}, 765 (2001)
\endreference

\reference{focuspipipi}
S.~Malvezzi,
arXiv:hep-ex/0307055, Contributed to the High Energy Physics Workshop on Scalar Mesons, 2003
\endreference

\reference{e791pipipi}
E.~M.~Aitala {\it et al.}  [E791 Collaboration],
Phys.\ Rev.\ Lett.\  {\bf 86}, 770 (2001)
\endreference

\reference{e791kpipi}
E.~M.~Aitala {\it et al.}  [E791 Collaboration],
Phys.\ Rev.\ Lett.\  {\bf 89}, 121801 (2002)
\endreference

\reference{mixing}
See the Note on $D{^0}\hbox{--}\overline D{^0}$ Mixing by D. Asner
on Page 675-679 of the Review of Particle Physics, Phys.Lett.B592: 1,2004; 
Dalitz plot analysis of the wrong sign rate
$D^0\to K^+\pi^-\pi^0$\cite{cleob} and the time dependence of Dalitz plot analysis of
$D^0\to K_S\pi^+\pi^-$\cite{cleoa} are two candidate processes
\endreference

\reference{theory}
M Bauer, B. Stech and M. Wirbel, Z. Phys. C {\bf 34}, 103 (1987);
P. Bedaque, A. Das and V.S. Mathur, Phys. Rev. D {\bf 49}, 269 (1994);
L.-L. Chau and H.-Y. Cheng, Phys. Rev. D {\bf 36}, 137 (1987);
K. Terasaki, Int. J. Mod. Phys. A {\bf 10}, 3207 (1995);
F. Buccella, M. Lusignoli and A. Pugliese, Phys. Lett. B {\bf 379}, 249 (1996)
\endreference

\reference{JDJackson}
J.~D.~Jackson, Nuovo Cim.\ {\bf 34}, 1644 (1964)
\endreference

\reference{dalitz}
R. H. Dalitz, {\it Phil. Mag.} {\bf 44}, 1068 (1953)
\endreference

\reference{Kin_Review}
See the review on 
Kinematics revised by J.~D.~Jackson 
on Page 675-679 of the Review of Particle Physics, Phys.Lett.B592: 1,2004;
\endreference


\reference{Kmatrix}
S.U. Chung {{\it et al.}}, Ann.\ Physik.\ {\bf 4}, 404 (1995)
\endreference

\reference{aitchison}
I.~J.~R.~Aitchison, Nucl.\ Phys.\ {\bf A 189}, 417 (1972)
\endreference

\reference{Blatt-Weisskopf}
J.~Blatt and V.~Weisskopf, {\it Theoretical Nuclear Physics}, New York: John
Wiley \& Sons (1952)
\endreference

\reference{Hippel-Quigg}
F.~von Hippel and C.~Quigg,
Phys.\ Rev.\ D {\bf 5}, 624, (1972)
\endreference

\reference{Zemach}
C. Zemach, Phys. Rev. B {\bf 133}, 1201 (1964);
C. Zemach, Phys. Rev. B {\bf 140}, 97 (1965)
\endreference


\reference{helicity}
S.~U.~Chung, Phys.\ Rev.\ D {\bf 48}, 1225, (1993);
J.~D.~Richman, CALT-68-1148
\endreference

\reference{pilkuhn} H.~Pilkuhn, {\it The Interactions of Hadrons},
Amsterdam: North-Holland (1967)
\endreference

\reference{flatte}
S.~M.~Flatte,
Phys.\ Lett.\ B {\bf 63}, 224 (1976).
\endreference

\reference{Kopp:2000gv}
S.~Kopp {\it et al.}  [CLEO Collaboration],
Phys.\ Rev.\ D {\bf 63}, 092001 (2001)
\endreference

\reference{GEANT}
R. Brun {\it et al.} GEANT 3.15, CERN Report No. DD/EE/84-1 (1987)
\endreference


\reference{markii}
R.~H.~Schindler {\it et al.} [Mark II Collaboration],
Phys. Rev. D{\bf 28}, 78 (1981)
\endreference

\reference{markiii}
J.~Adler {\it et al.} [Mark III Collaboration],
Phys. Lett. B{\bf  196}, 107 (1987)
\endreference

\reference{e687a}
P.L.~Frabetti {\it et al.} [E687 Collaboration],
Phys. Lett. B{\bf  286}, 195 (1992)
\endreference

\reference{e691} 
J.C.~Anjos {\it et al.} [E691 Collaboration],
Phys. Rev. D{\bf  48} 56 (1993)
\endreference

\reference{argus} 
H.~Albecht {\it et al.} [ARGUS Collaboration],
Phys. Lett. B{\bf  308}, 435 (1993)
\endreference

\reference{e687b} 
P.L.~Frabetti {\it et al.} [E687 Collaboration],
Phys. Lett. B{\bf  331}, 217 (1994)
\endreference

\reference{babardalitz}
B.~Aubert {\it et al.} [BABAR Collaboration],
arXiv:hep-ex/0207089. Contributed to the 31$^{st}$ International Conference
on High Energy Physics (ICHEP 2002)
\endreference

\reference{giri2003}
D.~Atwood and A.~Soni,
Phys.\ Rev.\ D {\bf 68}, 033003 (2003);
A. Giri \etal, \prD68,054018(2003);

D.~Atwood and A.~Soni,
arXiv:hep-ph/0312100
\endreference

\reference{e687dskkpi}
P.~L.~Frabetti {\it et al.} [E687 Collaboration],
Phys.\ Lett.\ B{\bf 351}, 591 (1995)
\endreference

\reference{focusds}
S.~Malvezzi,
AIP Conf.\ Proc.\  {\bf 549}, 569 (2002)
\endreference


\reference{isospin2}
W.~Hoogland {\it et al.},
Nucl.\ Phys.\ B {\bf 69}, 266 (1974)
\endreference

\reference{e791pm}
I.~Bediaga  [E791 Collaboration],
arXiv:hep-ex/0307008. Contributed to the High Energy Physics Workshop on Scalar Mesons, 2003
\endreference

\reference{lass}
D.~Aston {\it et al.} [LASS Collaboration],
Nucl.\ Phys.\ B {\bf 296}, 493 (1988)
\endreference

\reference{kappa}
C.~Gobel  [E791 Collaboration],
arXiv:hep-ex/0307003
\endreference

\reference{Scalar_Review}
See the Note on Scalar Mesons 
by S. Spanier and N. T\o rnqvist in 
this {\it Review\rm}
\endreference

\reference{wa76}
T.~A.~Armstrong {\it et al.} [WA76 Collaboration], Z.\ Phys.\ C{\bf  51}, 351 (1991)
\endreference

\reference{cleoacp}
D.~M.~Asner {\it et al.}  [CLEO Collaboration],
Phys.\ Rev.\ D {\bf 70}, 091101 (2004)
\endreference

\endreferencelist

\minihead{BBmix\_S042224 (lbl100.mini.stablemeson)}
\minititle{Charm Dalitz Plot Analysis Formalism and Results (Expanded RPP-2004 Version)}
\IndexPageno{dalitzm}
\miniauthor{Written November 2003 by D. Asner (University of Pittsburgh) }

\medskip
\noindent{\boldface\bfit Introduction:}
%
Charm meson decay dynamics have been studied extensively over the last
decade. 
Recent studies of multi-body decays of charm mesons probe a variety
of physics including doubly-Cabibbo suppressed decays\cite{cleoa}, 
searches for $CP$ violation\cite{cleoa}\cite{cleob}\cite{cleopipipi0}, the properties
of established light mesons\cite{e687pipipi}\cite{e791dspipipi}\cite{focuspipipi} and the properties of $\pi\pi$\cite{cleoa}\cite{focuspipipi}\cite{e791pipipi}
and $K\pi$\cite{e791kpipi} S-wave states.
Future studies could improve sensitivity to $D{^0}\hbox{--}\overline D{^0}$ mixing\cite{mixing}.

Weak nonleptonic decays of charm mesons are expected to proceed dominantly
through resonant two-body decays in several theoretical models\cite{theory};
see Ref.\cite{JDJackson} for a review of resonance phenomenology.
These amplitudes are calculated with the Dalitz plot analysis technique\cite{dalitz}, 
which uses
the mininum number of independent observable quantities. 
For three-body final states when the parent particle is a scalar,
the decay rate\cite{Kin_Review} is 
$$
\Gamma = \frac{1}{(2\pi)^3 32 \sqrt{s^3}}\left|\cal M\right|^2dm^2_{12}dm^2_{23},
\EQN{eqn:gamma}
$$
where $m_{ij}$ is the invariant mass of $i-j$ and the coefficient of the amplitude includes all kinematic factors.
The scatter plot in $m^2_{12}$ versus $m^2_{23}$ is called a Dalitz plot. 
If $\left|\cal M \right|^2$ is constant the allowed region of the plot will be populated
uniformly with events.
Any variation in $\left|\cal M \right|^2$ over the Dalitz plot
is due to dynamical rather than kinematical effects.


\medskip
\noindent{\boldface\bfit Formalism:}
The amplitude of the process, $D \rightarrow rc, r \rightarrow ab$, is
given by
$$
\RPPalign{
{\cal M}_r(L,m_{ab},m_{bc}) & = \sum_\lambda \left<ab|r_\lambda\right>T_r(m_{ab}) \left<cr_\lambda|D\right> 
\EQN{eqn:a} \cr
& = Z(L,\vec{p},\vec{q}) B^{D}_L(|\vec{p}|) B^{r}_L(|\vec{q}|) T_r(m_{ab}),
}
$$
where the sum is over the helicity states $\lambda$ of the intermediate resonance particle $r$, 
$a$ and $b$ are the daughter
particles of the
resonance $r$, $c$ is the spectator particle, 
$L$ is the orbital angular momentum between $r$ and $c$,
$\vec{p}$ is the momentum of $c$ in
the $r$ rest frame and $\vec{q}$ is momentum of 
$a$ in the $r$ rest frame,
$Z$ 
describes the angular distribution of final state particles, $B^{D}_L$ and $B^{r}_L$
are the barrier factors for the production of $r-c$ and $a-b$, respectively,
with angular momentum $L$ and 
$T_r$ is the dynamical function describing the resonance 
$r$. 
Usually the resonances are modeled with a
Breit-Wigner and the nonresonant contribution $D\to abc$ is parameterized as an S-wave with no variation in magnitude or phase across the Dalitz plot.
Some more recent analyses have used the $K$-matrix formalism\cite{Kmatrix} 
with the $P$-vector approximation\cite{aitchison} to describe the $\pi\pi$ S-wave.

\medskip
\noindent{\boldface\bfit Barrier Factor $B_L$:}
The maximum angular momentum $L$ in a strong decay is limited by the linear
momentum $\vec{q}$. Decay particles moving slowly with an impact parameter 
(meson radius) $r$ of 
order 1~fm have difficulty generating sufficient angular momentum to 
conserve the spin of the resonance. 
The Blatt-Weisskopf\cite{Blatt-Weisskopf}\cite{Hippel-Quigg} 
functions $B_L$, given in \Table{barrier},
weight the reaction amplitudes to account for this spin-dependent effect.
These functions are normalized to give $B_L=1$ for 
$z=(\left|{\vec q}\right|r)^2=1$.
Another common formulation $B^\prime_L$, also in \Table{barrier}, is 
normalized to give $B^\prime_L=1$ for $z=z_0=(\left|{\vec q_0}\right|r)^2$
where $q_0$ is the value of $q$ when $m_{ab}=m_r$. 

\midtable{barrier}
\Caption
Blatt-Weisskopf barrier factors.
\endCaption
\vglue -18pt
\centerline{
\vbox{
\halign{
\centertab{#}&\centertab{#}&\centertab{#}\cr
\tableheaddoublerule
$L$  & $B_L(q)$  & $B^\prime_L(q,q_0)$  \cr 
\tableheadsinglerule
0  & 1 & 1  \cr \cr
1 & $\sqrt{\frac{2z}{1+z}}$  & $\sqrt{\frac{1+z_0}{1+z}}$ \cr  \cr
2 & $\sqrt{\frac{13z^2}{(z\!-\!3)^2\!+\!9z}}$ &  $\sqrt{\frac{(z_0\!-\!3)^2\!+\!9z_0}{(z\!-\!3)^2\!+\!9z}}$ \cr  \cr
&  where $z = \left|{\vec q}\right|r$ & and $z_0= \left|{\vec q_0}\right|r$\cr 
\tableheaddoublerule
}}}
\endtable

\medskip
\noindent{\boldface\bfit Angular Distribution:}
$Z$ depends on the spin $L$ of resonance $r$. 
The Zemach formalism\cite{Zemach} describes the angular distributions
in terms of Legendre polynomials 
but is only valid for reactions where $a$, $b$ and $c$ 
are spin-0. 
For final state particles with non-zero spin, the helicity formalism
is required\cite{helicity}. The sum over helicity states yields
angular distributions that are proportional to
Legendre polynomials when transversality is enforced ($m^2$ = $m^2_{ab}$ rather than
$m^2$ = $m^2_r$ in the helicity sum denominator, see \Eq{transversality}).

\medskip
\noindent{\boldface\bfit Vector Intermediate States:}
The angular and barrier factors of \Eq{eqn:a} for vector
resonances are
$$
\RPPalign{
\bra{cr_\lambda}\!\!\ket{D} & = \left[f_+(P_D+P_c)_\mu
+ f_-(P_D-P_c)_\mu\right] g_{D}{\cal \epsilon}^\mu_\lambda, \cr
\bra{ab}\!\!\ket{r_\lambda} & = g_{r}{\cal \epsilon}^{*\nu}_\lambda(P_a-P_b)_\nu\,,
\EQN{vec}
}
$$
where $P_i$ is the four-momentum of particle $i$, ${\cal \epsilon}$ is the polarization vector associated with each decay vertex,
$f_+=f_+(m^2_{ab})$ and $f_-=f_-(m^2_{ab})$ are strong interaction form factors and $g_{D}$
and $g_{r}$ are strong interaction coupling constants.

We evaluate the sum over helicities $\lambda=\pm 1, 0$ as
$$
\sum_\lambda {\cal \epsilon}_\lambda^{*\mu} {\cal \epsilon}_\lambda^{\nu}
= -g^{\mu\nu} + \frac{q^\mu q^\nu}{m^2}.
\EQN{transversality}
$$
Transversality is enforced if we take the denominator of the second term
to be $m^2_{ab}$ rather than $m^2_r$. 
This enforces a vector current and ${\cal \epsilon}^\mu_\lambda q_\mu = 0$ by construction and so the $f_-(m^2_{ab})$ term does not
contribute. 
The product of the remaining three factors $f_+g_rg_D$, are
approximated as a constant $f_+(m^2_{ab}=m^2_r)$ and 
$g_{r,D}$ are the $L=1$ Blatt-Weisskopf factors given in \Table{barrier},
respectively.

The angular distribution for vector intermediate states is given by
$$
\RPPalign{
Z & = (P_D+P_c)_\mu(-g^{\mu\nu} + q^\mu q^\mu/m^2_{ab})(P_a - P_b)_\nu \cr
& = (m^2_{bc} - m^2_{ac}) + (m^2_D - m^2_c)(m^2_a - m^2_b)/m^2_{ab} \cr
& = -2{\vec p}.{\vec q},
\EQN{Z}
}
$$
where the Zemach form\cite{Zemach}, following the last equality, is only obtained
when transversality is enforced.
%

\medskip
\noindent{\boldface\bfit Scalar Intermediate States:}
There are no polarization vectors associated with the decay vertices so the process 
is either an S-wave decay 
or a D-wave decay. The S-wave decay has a uniform angular distribution
and the $L=0$ Blatt-Weisskopf factor is also constant.
The 
D-wave decay amplitude is described by
$$
\RPPalign{
& {\cal M}_r\!=\! 
\left[f_+(m^2_{ab})(P_D\!+\!P_c)_\mu\!+\!f_-(m^2_{ab})(P_D\!-\!P_c)_\mu\right]\!g_{D}\cr 
& T_r(m_{ab})\!\left[g_+(m^2_{ab})(P_a\!+\!P_b)^\mu\!+\!g_-(m^2_{ab})(P_a\!-\!P_b)^\mu\right]\!g_{r}.
\EQN{dscalar}
}
$$
The $D$-wave contribution is suppressed, due to small momenta, relative
to the $S$-wave and has been neglected is charm Dalitz plot analyses.

\medskip
\noindent{\boldface\bfit Tensor Intermediate States:}
The angular and barrier factors of \Eq{eqn:a} for tensor
resonances are
$$
\RPPalign{
\bra{cr_\lambda}\!\!\ket{D} & = T_1(m^2_{ab})(P_D+P_c)_\mu(P_D+P_c)_\nu\
 g_{D}{\cal \epsilon}^{\mu\nu}_\lambda\,, \cr
\bra{ab}\!\!\ket{r_\lambda} & =  T_2(m^2_{ab})(P_a-P_b)_\alpha(P_a-P_b)_\beta\
 g_{r}{\cal \epsilon}^{*\alpha\beta}_\lambda\,,
\EQN{tensor}
}
$$
where ${\cal \epsilon}$ are the polarization tensors associated with each decay vertex,
$T_1(m^2_{ab})$ and $T_2(m^2_{ab})$ are strong interaction form-factors and $g_{D}$
and $g_{r}$ are strong interaction coupling constants.

We evaluate the sum over helicities $\lambda=\pm 2, \pm 1, 0$~\cite{pilkuhn} as
$$
\sum_\lambda {\cal \epsilon}_\lambda^{*\mu\nu} {\cal \epsilon}_\lambda^{\alpha\beta}
 =\frac{\left(\aleph^{\mu\alpha}\aleph^{\nu\beta}\!+\!
                  \aleph^{\mu\beta}\aleph^{\nu\alpha}\right)}{2}
\!-\!\frac{\aleph^{\mu\nu}\aleph^{\alpha\beta}}{3} 
\EQN{aleph}
$$
where $\aleph^{\mu\nu}\!=\!-g^{\mu\nu}\!+\!\frac{q_\mu q_\nu}{m^2}$.
Transversality is enforced if we take the denominator of the second term
to be $m^2_{ab}$ rather than $m^2_r$.
This enforces a tensor current and $q_{ab}^\mu{\cal \epsilon}_{\mu\nu} = q_{ab}^\nu{\cal \epsilon}_{\mu\nu} = 0$ by construction.
The product of the remaining four factors $T_1 T_2 g_rg_D$, are
approximated as a constant $T_{1,2}(m^2_{ab}=m^2_r)$ and
$g_{r,D}$ are the $L=2$ Blatt-Weisskopf factors given in \Table{barrier},
respectively.

The angular distribution for tensor intermediate states is
$$
\RPPalign{
Z  = & (P_D\!+\!P_c)_\mu\!(P_D\!+\!P_c)_\nu\!
\left[\frac{\left(\aleph^{\mu\alpha}\aleph^{\nu\beta}\!+\!
                  \aleph^{\mu\beta}\aleph^{\nu\alpha}\right)}{2}
\!-\!\frac{\aleph^{\mu\nu}\aleph^{\alpha\beta}}{3}\right] \cr
& (P_a - P_b)_\alpha(P_a - P_b)_\beta \cr
 = & \left(m^2_{bc}\!-\!m^2_{ac}\! 
+\!\frac{(m^2_D\!-\!m^2_c)(m^2_a\!-\!m^2_b)}{m^2_{ab}}\right)^2 \cr
& \!-\!\frac{1}{3}\left(m^2_{ab}\!-\!2m^2_D\!-\!2m^2_c\!+\!\frac{(m^2_D\!-\!m^2_c)^2}{m^2_{ac}}\right) \cr
& \!\times\!\left(m^2_{ab}\!-\!2m^2_a\!-\!2m^2_b\!+\!\frac{(m^2_a\!-\!m^2_b)^2}{m^2_{ac}}\right)\cr
 = & \frac{4}{3}\left[3\left({\vec p}.{\vec q}\right)^2 - \left(|{\vec p}||{\vec q}|\right)^2\right]
\EQN{Zten}
}
$$
where the Zemach form\cite{Zemach}, following the last equality, is only obtained
when transversality is enforced.

\medskip
\noindent{\boldface\bfit Dynamical Function $T_r$:}
The dynamical function $T_r$ is derived from the $S$-matrix formalism. 
In general, the amplitude that a final state $f$ couples to an initial state $i$
through the unitary scattering operator $S$,
$S_{fi} = \left<f|S|i\right>$.
where the scattering operator $S$ is unitary.
 and satisfies $SS^\dagger=S^\dagger S=I$. 
The transition operator 
${\hat T}$ is defined by separating the probability that $f=i$ yielding,
%
$$
S = I + 2iT = I + 2i\left\{\rho\right\}^{1/2}{\hat T}\left\{\rho\right\}^{1/2},
\EQN{eqn:s}
$$ 
where $I$ is the identity operator, ${\hat T}$ is Lorentz invariant transition operator,
$\rho$ is the diagonal phase space matrix where $\rho_{ii}=2q_i/m$ and $q_i$ is the
breakup momentum for decay channel $i$.
In the single channel S-wave scenario
$S=e^{2i\delta}$ 
satisfies unitarity and implies 
$$
{\hat T}=\frac{1}{\rho}e^{i\delta}\sin\delta.
\EQN{eqn:t}
$$ 

There are three common formulations of the dynamical function. The
Breit-Wigner formalism
is the simplest formulation.
 - the first term in a Taylor expansion about a $T$ matrix pole,
The $K$-matrix formalism\cite{Kmatrix} is more general, 
allowing more than one $T$ matrix pole
and coupled channels while preserving unitarity. The Flatte distribution\cite{flatte} is used to parameterize resonances near threshold and is equivalent to a one-pole,
two-channel $K$-matrix.

\medskip
\noindent{\boldface\bfit Breit-Wigner Formulation:}
The common formulation of the Breit-Wigner decaying to spin-0 particles 
$a$ and $b$ is
$$
T_r(m_{ab}) = \frac{1}{m^2_r - m^2_{ab} - im_r \Gamma_{ab}(q)}
\EQN{eqn:bw}
$$
where the ``mass dependent'' width $\Gamma_{ab}$ is

$$
	\Gamma_{ab} = \Gamma_r \left(\frac{q}{q_0}\right)^{2L+1}
\left(\frac{m_0}{m_{ab}} \right) B^\prime_L(q,q_0)^2.
\EQN{eqn:gammaab}
$$
A Breit-Wigner parametrization best describes isolated, non-overlapping resonances far from the threshold of additional decay channels. 
Unitarity can be violated when the dynamical function is parameterized as the sum
of two or more overlapping Breit-Wigners.
The proximity of a threshold to the resonance shape distorts the line
shape from a simple Breit-Wigner. 
This scenario is described by the Flatte formula and is discussed below.

\medskip
\noindent{\boldface\bfit $K$-matrix Formalism:}
The $T$ matrix can be described as
$$
{\hat T} = (I - i{\hat K}\rho)^{-1} {\hat K},
\EQN{eqn:tmatrix}
$$
where 
${\hat K}$ is the Lorentz invariant
$K$-matrix describing the 
scattering process and 
$\rho$ is the phase space factor.

Resonances appear as a sum of poles in the $K$-matrix
$$
{\hat K}_{ij} = \sum_\alpha \frac{m_\alpha \Gamma_{\alpha i}(m) m_\alpha \Gamma_{\alpha j}(m) }
{(m^2_\alpha - m^2)\sqrt{\rho_i\rho_j}}.
$$
For the special case of a single channel, single pole we obtain
$$
K = \frac{m_0\Gamma(m)}{m^2_0 - m^2}
$$
and
$$
T = K(1-iK)^{-1} =  \frac{m_0\Gamma(m)}{m^2_0 - m^2 -im_0\Gamma(m)}
$$
which is the relativistic Breit-Wigner formula.
For the special case of a single channel, two poles we have
$$
K = \frac{m_\alpha\Gamma_\alpha(m)}{m^2_\alpha - m^2} 
+ \frac{m_\beta\Gamma_\beta(m)}{m^2_\beta - m^2}
$$
and in the limit that $m_\alpha$ and $m_\beta$ are far apart relative to the widths we
can approximate the $T$ matrix as the sum of two Breit-Wigners,
$T(K_\alpha+K_\beta) \approx T(K_\alpha) + T(K_\beta)$,
$$
T \approx  \frac{m_\alpha\Gamma_\alpha(m)}{m^2_\alpha - m^2 -im_\alpha\Gamma_\alpha(m)} 
+  \frac{m_\beta\Gamma_\beta(m)}{m^2_\beta - m^2 -im_\beta\Gamma_\beta(m)}.
\EQN{eqn:t2matrix}
$$
In the case of two nearby resonance \Eq{eqn:t2matrix} is not valid and exceeds unity 
(violates unitarity).

This formulation, 
which applies to s-channel production in two-body scattering
$ab \to cd$, 
can be generalized to describe the production of resonances
in processes, such as the decay of charm mesons. The key assumption here
is that the two-body system described by the $K$-matrix does {\it not} interact
with the rest of the final state\cite{aitchison}. The quality of this assumption varies with 
the production process and is appropriate for scattering experiments like $\pi^- p \to \pi^0\pi^0 n$, radiative decays such as $\phi, J/\psi \to \gamma \pi\pi$ and semileptonic decays such as $D \to K\pi\ell\nu$. This assumption may be of limited validity
for production processes such as $p{\overline p} \to \pi\pi\pi$ or $D \to \pi\pi\pi$.
In these scenarios the two-body Lorentz invariant amplitude, ${\hat F}$, is given as
$$
{\hat F}_i = (I - i {\hat K}\rho)_{ij}^{-1} {\hat P}_j = {\hat T}{\hat K}^{-1}{\hat P}
\EQN{eqn:tmatrix}
$$
where 
$P$ is the production vector which parameterizes the resonance
production in the open channels. 

For the $\pi\pi$ S-wave
the five channels relevant for $D$ decays,
are $\pi\pi$, $KK$, $\eta\eta$, $\eta^\prime\eta^\prime$ and $4\pi$. 
In this scenario
a common formulation of the $K\!-\!matrix$ is
$$
K_{ij}\!(s)\!=\!\left\{ \sum_\alpha \frac{g_i^{(\alpha)}g_j^{(\alpha)}}{m_\alpha^2\!-\!s}\!+\!f_{ij}^{sc}\frac{1 \!-\! s_0^{sc}}{s\!-\!s_0^{sc}}\right\}
\times\frac{s\!-\!s_A/2m^2_{\pi}}{(s\!-\!s_{A0})(1\!-\!s_{A0})}.
\EQN{eqn:kmatrix}
$$
The factor $g_i^{(\alpha)}$ is the coupling constant of the $K$-matrix pole
$m_\alpha$ to meson channel $i$; the parameters $f_{ij}^{sc}$ and $s_0^{sc}$
describe a smooth part of the $K$-matrix elements; the multiplicative factor 
$\frac{s-s_A/2m_\pi^2}{(s\!-\!s_{A0})(1\!-\!s_{A0})}$ 
suppresses a false
kinematical singularity near the $\pi\pi$ threshold - the Adler zero; and 
the number 1 has units ${\rm GeV}^2$.

The production vector, with $i=1$ denoting $\pi\pi$, is 
$$
P_{j}\!(s) = \left\{ \sum_\alpha \frac{\beta_\alpha g_j^{(\alpha)}}{m_\alpha^2\!-\!s}+ f_{1j}^{pr}\frac{1 \!-\! s_0^{pr}}{s\!-\!s_0^{pr}}\right\}
 \times  \frac{s\!-\!s_A/2m^2_{\pi}}{(s\!-\!s_{A0})(1\!-\!s_{A0})}\,,
\EQN{eqn:pvector}
$$
where the free parameters of the Dalitz plot fit are the production 
coupling $\beta_\alpha$,
and the production vector background parameters $f_{1j}^{pr}$ and $s_0^{pr}$.
All other parameters are fixed by scattering experiments.

\medskip
\noindent{\boldface\bfit Flatte Formalism:}
The scenario where another channel opens close to the resonance position is described by the Flatte formulation
$$
{\hat T(m_{ab})}\!=\!\frac{{\hat g}}{m^2_r-m_{ab}^2-i(\rho_1g_1^2+\rho_2g_2^2)},\,\,\hfill\,\,g_1^2+g_2^2 = m_r\Gamma_r.
\EQN{eqn:flatte}
$$
This situation occurs in the $\pi\pi$ S-wave where the $f_0(980)$ is near the $K{\overline K}$ threshold and in the $\pi\eta$ channel where the $a_0(980)$ also lies near $K{\overline K}$ threshold. For the $a_0(980)^+$ resonance 
the relevant coupling constants are $g_1=g_{\pi\eta}$ and $g_2 = g_{KK}$.
For the $f_0(980)$ the relevant coupling constants are 
$g_1 = g_{\pi\pi}$ and $g_2 = g_{KK}$ where the charged
and neutral $K$ channels are usually assumed to have the same coupling constant but
separate phase space factors due to $m_{K^+} \neq m_{K^0}$.

\medskip
\noindent{\boldface\bfit Branching Ratios from Dalitz Plot Fits:}
The fit to the Dalitz plot distribution using either the Breit-Wigner or the
$K$-matrix formalism factorizes into a resonant contribution to the amplitude ${\cal M}_j$
and a complex coefficient, $a_j e^{i\delta_j}$, 
where $a_j$ and $\delta_j$ are real. 
The definition of a rate of a single process, given a set of
amplitudes $a_j$ and phases $\delta_j$ is the square of 
the relevant matrix
element. In this spirit, the fit fraction is usually defined as
the integral over the Dalitz plot ($m_{ab}$ vs $m_{bc}$) 
of a single amplitude
squared divided by the integral over the Dalitz plot
of the square of the coherent sum of all amplitudes,
$$
{\rm Fit\, Fraction}_j =  \frac{\int \left| 
a_j e^{i\delta_j}{\cal M}_j
\right|^2 dm^2_{ab}dm^2_{bc}}
{ \int \left| \sum_k  
a_k e^{i\delta_k}
{\cal M}_k \right|^2
dm^2_{ab}dm^2_{bc}}
\EQN{eqn:FF}
$$
where ${\cal M}_j$ is defined by \Eq{eqn:a} and 
described in Ref.\cite{Kopp:2000gv}.
The sum of the fit fractions for all components
will in general not be unity due to interference.

\medskip
\noindent{\boldface\bfit Reconstruction Efficiency and Resolution:}
The efficiency for reconstructing an event as a function of position
on the Dalitz plot is in general non-uniform. 
Typically, a signal Monte Carlo sample, utilizing full
GEANT\cite{GEANT} detector simulation generated with a uniform 
distribution in phase
space is used to determine the efficiency.
The variation in efficiency across the Dalitz plot varies with experiment and decay mode.

Finite detector resolution can usually be safely neglected as most resonances are
comparatively broad. Notable exceptions where detector resolution effects must be
modeled are $\phi \to K^+K^-$ and $\omega \to \pi^+\pi^-$. Additionally, the momenta
of $a$,$b$ and $c$ can recalculated with a $D$ mass constraint. This forces the 
kinematical boundaries of the Dalitz plot to be strictly respected. 

\medskip
\noindent{\boldface\bfit Background Parametrization:}
The contribution of background to the charm samples varies by
experiment and final state. 
The background naturally falls into four categories:
(i) purely combinatoric background containing no resonances,
(ii) combinatoric background containing intermediate resonances, such as a real $K^{*-}$
or $\rho$, plus additional random particles,
(iii) mistagged decays such as a real $\overline{D}^0$ incorrectly identified as $D^0$ and
(iv) misidentified $D$ daughters such as
 $D^+\to \pi^-\pi^+\pi^+$ or $D_s^+\to K^-K^+\pi^+$ reconstructed as
 $D^+\to K^-\pi^+\pi^+$.


The contribution from combinatoric background resonances is distinct 
from the 
resonances in the signal because the former do $not$ interfere with the
latter 
since they are not from true $D$'s. 
The usual identification tag of the initial particle as a
$D{^0}$ or a $\overline D{^0}$
is the charge of the distinctive slow
pion in the decay sequence $D^{*+}\!\to\!D{^0}\pi^+$ or
$D^{*-} \to \overline D^0 \pi^-$.  
Another possibility
is the identification of one of the $D$'s from
$\psi(3770)\!\to\!D^0\overline D{^0}$.
The mistagged background is subtle and may be mistakenly enumerated in the $signal$ fraction
determined by a $D^0$ mass fit. Mistagged decays
contain true $\overline{D}^0$'s and so 
the resonances in the mistagged sample exhibit interference 
on the Dalitz plot. 
Background from mis-identified $D$ daughters - if present -
is the most problematic. 

%

\medskip
\noindent{\boldface\bfit Experimental Results:}
The Dalitz plot analysis technique has been applied to the decays $D \to rc$, $r\to ab$
where the decay products $a$, $b$ and $c$ are $K$ or $\pi$ and the intermediate state $r$ is a scalar, vector or tensor meson. More generally the decay products could also be the pseudo-scalar $\eta$ or $\eta^\prime$ mesons or narrow vector 
$\omega$ or $\phi$ mesons.
The set of charm Dalitz plot analyses reported by experiments are listed in 
\Table{Dalitz}. 

\midtable{Dalitz}
\Caption
Reported Dalitz Plot Analyses.
\endCaption
\vglue -18pt
\centerline{
\vbox{
\halign{
\lefttab{#}&\centertab{#}\cr
\tableheaddoublerule
Decay  & Experiment(s) \cr 
\tableheadsinglerule
$D^0 \to K_S\pi^+\pi^-$   & Mark~II$\!$\cite{markii},Mark~III$\!$\cite{markiii} \cr
                          & E691$\!$\cite{e691},E687$\!$\cite{e687a}\cite{e687b}\cr
                          & ARGUS$\!$\cite{argus}, CLEO$\!$\cite{cleoa}  \cr
$D^0 \to K^-\pi^+\pi^0$   & Mark~III$\!$\cite{markiii},E687$\!$\cite{e687b}  \cr
                          & E691$\!$\cite{e691},CLEO$\!$\cite{Kopp:2000gv} \cr
$D^0 \to {\overline K}^0K^+\pi^-$ & BABAR$\!$\cite{babardalitz}  \cr 
$D^0 \to K^0 K^-\pi^+$    & BABAR$\!$\cite{babardalitz}  \cr
$D^0 \to \pi^+\pi^-\pi^0$ & CLEO$\!$\cite{cleopipipi0}  \cr
$D^0 \to K_S K^+K^-$      & BABAR$\!$\cite{babardalitz} \cr
$D^+ \to K^-\pi^+\pi^+$   & Mark~III$\!$\cite{markiii},E687$\!$\cite{e687b} \cr
                          & E691$\!$\cite{e691},E791$\!$\cite{e791kpipi} \cr
$D^+ \to {\overline K}^0\pi^+\pi^0$ & Mark~III$\!$\cite{markiii} \cr
$D^+ \to \pi^+\pi^+\pi^-$ & E687$\!$\cite{e687pipipi},E791$\!$\cite{e791pipipi},FOCUS$\!$\cite{focuspipipi}\cr
$D^+ \to K^+K^-\pi^+$ & E687$\!$\cite{e687dskkpi},FOCUS$\!$\cite{focusds} \cr 
$D_s^+ \to K^+K^-\pi^+$ & E687$\!$\cite{e687dskkpi},FOCUS$\!$\cite{focusds}\cr
$D_s^+ \to \pi^+\pi^+\pi^-$ & E687$\!$\cite{e687pipipi},E791$\!$\cite{e791dspipipi},FOCUS$\!$\cite{focuspipipi} \cr
\tableheaddoublerule
}}}
\endtable

\medskip
\noindent{\boldface\bfit $D{^0}\to K^0_S\pi^+\pi^-$ ---}
Several experiments have analyzed the decay $D{^0}\to K^0_S\pi^+\pi^-$.
The earliest analyses, by Mark~II\cite{markii}, Mark~III\cite{markiii}, 
and E687\cite{e687a}, assumed only two intermediate resonances, 
$K^0_S \rho^0$, $K^*(892)^-\pi^+$, and a significant nonresonant component. 
Additional resonances were considered by 
E691\cite{e691} but were not found to be statistically significant. 
ARGUS\cite{argus} and E687\cite{e687b}, with more events, 
fit the Dalitz plot with six intermediate resonances:
$K^{*}(892)^- \pi^+$, ${K}_{0}^{*}(1430)^- \pi^+$, $K^0_S \rho^0$, 
$K^0_S f_0(975)$, $K^0_S f_2(1270)$, and $K^0_S f_0(1400)$. 
The nonresonant contribution was negligible. 
The early and later E687 results [\use{Ref.e687a},~\use{Ref.e687b}] 
were consistent under similar assumptions.
The most precise results are from CLEO\cite{cleoa}, which includes three
additional resonances: $K^0_S \omega, K^{*}(1680)^- \pi^+$ and the doubly 
Cabibbo-suppressed $K^{*}(892)^+ \pi^-$. 
They find a much smaller nonresonant contribution than did the earliest experiments.

It is not straightforward to compare or
combine results using different descriptions 
of the angular distributions, barrier factors, resonant parametrizations,
and different sets of resonances.
Some of the earlier results 
\citerange{markiii}{e691}, 
did not include barrier factors [\use{Ref.Blatt-Weisskopf},~\use{Ref.Hippel-Quigg}].
Most of the earlier results 
[\use{Ref.markiii}--\use{Ref.e691},~\use{Ref.e687b}] 
used the Zemach formalism\cite{Zemach} to describe the angular shape
of the decay pattern, while the more recent results [\use{Ref.argus},~\use{Ref.cleoa}]
use the helicity formalism\cite{helicity}.

The significance of the nonresonant component in the smaller data
samples has been attributed to the presence of the broad scalar resonances 
$K^*_0(1430)^-$ and $f_0(1370)$ that were later observed in the larger data 
samples. The observation of a small but significant nonresonant 
component in the largest data samples suggests the presence of 
additional broad scalar resonances, the $\kappa(800)$ and $\sigma(500)$. The CLEO analysis
could accomodate the $\sigma(500)$ in lieu of the nonresonant component, but found no evidence
for the $\kappa(800)$.

\medskip
\noindent{\boldface\bfit $D{^0}\to \pi^+\pi^-\pi^0$ 
  {\rm \boldface and} $D{^0}\to {\overline K^0}K^+K^-$ ---}
The only significant contribution to the resonant substructure 
of $D{^0}\to \pi^+\pi^-\pi^0$ is in the $\rho\pi$ channels. 
A small nonresonant component is observed but all other $\pi\pi$ resonances, 
including the $\sigma(500)$, yielded fit fractions consistent with zero. 
The CLEO\cite{cleopipipi0} results for $D{^0}\to \pi^+\pi^-\pi^0$ are
given in \Table{pipipi0}.

\midtable{pipipi0}
\Caption
Dalitz fit results of $D{^0}\to \pi^+\pi^-\pi^0$. 
\endCaption
\vglue -18pt
\centerline{
\vbox{
\halign{
\centertab{#}&\centertab{#}&\centertab{#}&\centertab{#}\cr
\tableheaddoublerule
Resonance & Amplitude & Phase($^\circ$) & Fit fraction(\%) \cr 
\tableheadsinglerule
$\rho^+$ & $1.$ (fixed) & $0.$ (fixed) & $76.5 \pm 1.8 \pm 4.8$ \cr
$\rho^0$ & $0.56\pm 0.02 \pm 0.07$ & $10 \pm 3 \pm 3$ & $23.9 \pm 1.8 \pm 4.6$ \cr
$\rho^-$ & $0.65 \pm 0.03 \pm 0.04$ & $-4 \pm 3 \pm 4$ & $32.3 \pm 2.1 \pm 2.2$ \cr 
nonresonant & $1.03 \pm 0.17 \pm 0.31$ & $77 \pm 8 \pm 11$ & $2.7 \pm 0.9 \pm 1.7$ \cr 
\tableheaddoublerule
}}
}
\endtable

The BABAR\cite{babardalitz} results for $D{^0}\to {\overline K^0}K^+K^-$
are given in \Table{kkk}. The non-$\phi$ resonant substructure in $K^+K^-$ is significant. 
Resonant constributions from $a_0(980)^0$, $a_0(980)^+$, and $f_0(980)$ are observed. 
The nonresonant and the doubly Cabibbo-suppressed contributions are 
consistent with zero.

\midtable{kkk}
\Caption
Dalitz fit results of $D{^0}\to {\overline K^0}K^+K^-$. 
\endCaption
\vglue -18pt
\centerline{
\vbox{
\halign{
\lefttab{#}&\lefttab{#}&\lefttab{#}\cr
\tableheaddoublerule
Resonance\hglue 0.35in        & Phase($^\circ$)  \hglue 0.3in     & Fit fraction(\%) \cr
\tableheadsinglerule
${\overline K^0}\phi$     & $0.$ (fixed)    & $45.4\pm 1.6 \pm 1.0$ \cr
${\overline K^0}a_0(980)$ & $109 \pm 5$     & $ 60.9 \pm 7.5 \pm 13.3$ \cr
${\overline K^0}f_0(980)$ & $-161 \pm 14$   & $12.2 \pm 3.1 \pm 8.6$ \cr
$a_0(980)^+K^-$  & $-53\pm 4$   & $34.3 \pm 3.2 \pm 6.8$   \cr
$a_0(980)^-K^+$  & $-13\pm 15$  & $3.2 \pm 1.9 \pm 0.5 $   \cr
nonresonant      & $ 40 \pm 44$ & $ 0.4 \pm 0.3 \pm 0.8$   \cr
\tableheaddoublerule
}}
}
\endtable

Charm Dalitz plot analyses might be useful for calibrating tools used
in $B$ decays: specifically, to extract $\alpha$ from $B{^0}\to \pi^+\pi^-\pi^0$,
$\beta$ from $B{^0}\to {\overline K^0}K^+K^-$, and $\gamma$ from 
$B{^\pm}\to D K^\pm$ followed by $D \to {\overline K^0} K^+ K^-$ or
$D \to {\overline K^0} \pi^+ \pi^-$~\cite{giri2003} .

\medskip
\noindent{\boldface\bfit $D^+ \to \pi^+ \pi^+ \pi^-$: 
    a $\sigma(500)$ or $f_0(600)$ ---}
The decay $D{^+}\to\pi^+\pi^+\pi^-$
has been studied by the E687\cite{e687pipipi}, E791\cite{e791pipipi} and FOCUS\cite{focuspipipi} 
experiments. The E687 experiment considered the $\rho(770)^0 \pi^+$, $f_0(980) \pi^+$, 
$f_2(1270) \pi^+$, and a nonresonant component. The E791 experiment included
in addition $f_0(1370) \pi^+$ and $\rho(1450)^0 \pi^+$. 
Both analyses found a very large fraction $(\sim50\%)$ for the nonresonant
contribution, perhaps indicating a broad scalar contribution. 
E791 found the nonresonant amplitude to be consistent with zero if
a broad scalar resonance was included in the fit.
FOCUS analyzed its data sample using both the Breit-Wigner 
formalism and the $K$-matrix formalism. 
The Breit-Wigner analysis included
$\rho(770), f_0(980), f_2(1270), f_0(1500), \sigma(500)$, and a nonresonant contribution. 
Applying the $K$-matrix formalism to the S wave and parameterizing the 
$\rho(770)$ and $f_2(1270)$ with the Breit-Wigner functions also described the
FOCUS data well. 

None of these analyses has modeled the dynamics of the $\pi^+\pi^+$ interaction. 
Consideration of the $I=2$ S-wave and D-wave phase shifts, also
measured in $\pi^+ p \to \pi^+\pi^+ n$\cite{isospin2}, could affect the
$\pi^+\pi^-$ S-wave result.

E791 finds additional evidence that the low mass $\pi\pi$ feature 
is resonant by examining the phase
of the $\pi\pi$ amplitude in the vicinity of the reported 
$\sigma(500)$ mass. A phase variation with invariant $\pi\pi$ mass is
consistent with a resonant contribution \cite{e791pm}.

\Table{sigma500} gives the parameters of
the $\sigma(500)$ determined in charm Dalitz plot analyses. A consistent relative 
phase between the $\sigma(500)$ and $\rho(770)$ resonances is observed.

\midtable{sigma500}
\Caption
Parameters of the $\sigma(500)$ resonance. The amplitude and phase
are relative to the $\rho(770)$. 
\endCaption
\vglue -18pt
\centerline{
\vbox{
\halign{
\lefttab{#}&\centertab{#}&\centertab{#}&\centertab{#}\cr
\tableheaddoublerule
Experiment             & E791\cite{e791pipipi}   & CLEO\cite{cleoa}        & FOCUS\cite{focuspipipi} \cr
\tableheadsinglerule
Decay Mode             & $D^+\to\pi^+\pi^+\pi^-$ &$D^0\to K^0_S\pi^+\pi^-$ &$D^+\to\pi^+\pi^+\pi^-$  \cr
Amplitude              & $1.17\pm0.13\pm0.06$    & $0.57\pm0.13$           &  ---                    \cr
Phase($^\circ$)        & $205.7\pm 8.0\pm 5.2$   & $214\pm 11$             & $200\pm 31$             \cr
$m({\rm MeV}/c^2)$     & $478^{+24}_{-23}\pm 17$ & $513\pm 32$             & $443\pm 27$             \cr
$\Gamma({\rm MeV}/c^2)$& $324^{+42}_{-40}\pm 21$ & $335\pm 67$             & $443\pm 80$             \cr
\tableheaddoublerule
}}
}
\endtable

\medskip
\noindent{\boldface\bfit $D^+ \to K^- \pi^+ \pi^+$: a $\kappa(800)$? ---}
Indication of a broad $K\pi$ scalar intermediate resonance has
been reported by E791 in the decay $D^+ \to K^-\pi^+\pi^+$\cite{e791kpipi}.
Fitting the Dalitz plot with 
${\overline K}^*(892)^0 \pi^+$, ${\overline K}^*_0(1430)^0 \pi^+$,
${\overline K}^*_2(1430)^0 \pi^+$, and ${\overline K}^*(1680)^0 \pi^+$, 
plus a constant nonresonant component,
E791 finds results consistent with earlier results 
from E691 and E687 with a nonresonant fit fraction of over 90\%.
Having reconstructed more events than the other experiments, 
E791 was led to include an extra 
low-mass S-wave $\overline K \pi$ 
resonance to account for the poor fit already seen by earlier experiments: 
A $\kappa(800)$ with ${\rm} m = 797\pm 19\pm 43$~MeV/$c^2$ 
and $\Gamma = 410\pm 43\pm 87$~MeV/$c^2$ much improved the fits.
The $\kappa(800)$ is now the dominant resonance and the nonresonant 
fit fraction is reduced from $90.9\pm 2.6\%$ to $13.0\pm 5.8\pm 4.4\%$.
As discussed with the $\sigma(500)$, the $K^-\pi^+$ S-wave result could
be affected by modeling the dynamics of the $I=2$ $\pi^+\pi^+$ interaction.

E791 also modeled the $K\pi$ S-wave phase variation as a function of $K\pi$
mass with the $K^*_0(1430)$ resonance only and a nonresonant component 
following the parameterization of LASS\cite{lass}. It was necessary to
relax the unitarity constraint to describe 
the E791 data\cite{kappa}. 
The $K\pi$ S-wave 
phase behavior in this model is consistent with the model that includes 
the $\kappa$ resonance.

CLEO allowed scalar $K\pi$ resonances in the fit to
$D^0 \to K^-\pi^+\pi^0$\cite{Kopp:2000gv} and $D^0 \to K^0_S\pi^+\pi^-$\cite{cleoa} 
and observed a significant contribution for only $K^*_0(1430)$\cite{Scalar_Review}. 
BABAR has analyzed the decay $D^0\to K^0K^-\pi^+$ and 
$D^0\to {\overline K}^0K^+\pi^-$\cite{babardalitz}.
They fit the former Dalitz plot with both positively charged and 
neutral ${\overline K}^*(892)$, ${\overline K}^*_0(1430)$,
${\overline K}^*_2(1430)$, ${\overline K}^*(1680)$ and
 $a_0(980)^-$,  $a_0(1450)^-$, $a_2(1310)^-$ resonances, and a nonresonant component. 
The second Dalitz plot is fit with the identical resonances except for the $a_2(1310)^-$.
A good fit is obtained in both cases without including the $\kappa$.

\medskip
\noindent{\boldface\bfit $f_0(980)$, $f_0(1370)$ {\rm \boldface and} $f_0(1500)$ ---}
The proximity of the $K \overline{K}$ threshold requires a coupled-channel 
or Flatte parametrization\cite{flatte} of the $f_0(980)$ in charm Dalitz plot analyses.
The width of the $f_0(980)$ is poorly known.  
E791 used a coupled-channel Breit-Wigner function, 
following the parametrization of Ref.\cite{wa76},
to describe the $f_0(980)$ in $D^+_s \to \pi^+\pi^+\pi^-$ \cite{e791dspipipi}, 
and measured $m_r=977\pm3\pm 2$ MeV/$c^2$, $g_{\pi\pi}=0.09\pm 0.01\pm 0.01$,
and $g_{KK} = 0.02\pm 0.04\pm 0.03$. 
Results similar to these are desirable for input to
the analysis of the $D_s^+ \to K^+K^-\pi^+$\cite{focusds},
which includes the $f_0(980)$ and $a_0(980)$.

The quark content of the $f_0(1370)$ and $f_0(1500)$ can perhaps 
be inferred from how they populate various Dalitz plots. The E791 analysis
of $D^+ \to \pi^+\pi^+\pi^-$\cite{e791pipipi} finds a contribution from 
the $f_0(1370)$ but not the $f_0(1500)$.
The FOCUS analysis\cite{focuspipipi} of this decay does not find a 
significant contribution from the $f_0(1370)$.
For the $D^+_s \to \pi^+\pi^+\pi^-$, E687\cite{e687pipipi} 
and FOCUS\cite{focuspipipi} do not see the $f_0(1370)$ but do see a
resonance with parameters similar to the $f_0(1500)$,
while E791\cite{e791dspipipi} observes a $\pi\pi$ resonance 
($m = 1434\pm18 \pm 9$~MeV/$c^2$ and $\Gamma = 172\pm 32\pm 6$~MeV/$c^2$) 
that is not consistent with either meson. 
BABAR has found no evidence for either the $f_0(1370)$ or the $f_0(1500)$ in
$D^0 \to \overline K^0 K^+ K^-$\cite{babardalitz}, 
while CLEO has observed the $f_0(1370)$ in
$D^0 \to K^0_S \pi^+\pi^-$\cite{cleoa}. 
Future analyses will present a clearer picture only if the same resonances 
and model of decay amplitudes are applied to all Dalitz plot fits.

\medskip
\noindent{\boldface\bfit Doubly Cabibbo-Suppressed Decays ---}
There are two classes of multibody doubly Cabibbo-suppressed (DCS) decays of 
charm mesons. 
The first consists of those in which the DCS and corresponding 
Cabbibo-favored (CF) decays populate distinct Dalitz plots:
the pairs $D^0 \to K^+\pi^-\pi^0$ and $D^0 \to K^-\pi^+\pi^0$, 
or $D^+ \to K^+\pi^+\pi^-$ and $D^+ \to K^-\pi^+\pi^+$, are examples.
CLEO\cite{cleob} has reported $\frac{{\cal B}(D^0 \to K^+\pi^- \pi^0)}{{\cal B}(D^0 \to
K^- \pi^+\pi^0)} = (0.43^{+0.11}_{-0.10}\pm 0.07)\%$.

The second class consists of decays where the 
DCS and CF modes populate the same Dalitz plot: for example, 
$D^0 \to K^{*-} \pi^+$ and $D^0 \to K^{*+} \pi^-$
both contribute to $D^0 \to K^0_S \pi^+\pi^-$. 
In this case, the potential for interference of DCS and CF amplitudes 
increases the sensitivity to the DCS amplitude. 
CLEO\cite{cleoa} has reported the relative amplitudes and phases to be
$(7.1 \pm 1.3^{+2.6\,+2.6}_{-0.6\,-0.6})\%$ and $(189\pm 10\pm 3^{+15}_{-5})^\circ$, 
respectively, corresponding to
$\frac{{\cal B}(D^0 \to K^\ast(892)^+ \pi^-)}{{\cal B}(D^0 \to
K^\ast(892)^- \pi^+)} = (0.5 \pm 0.2 ^{+0.5}_{-0.1}\,^{+0.4}_{-0.1})\%$.

\medskip
\noindent{\boldface\bfit $CP$ Violation ---}
In the limit of $CP$ conservation, charge conjugate decays will have the same
Dalitz plot distribution. 
The $D^{*\pm}$ tag enables the discrimination between $D^0$ and ${\overline D}^0$. 
The integrated $CP$ violation across the Dalitz plot is determined from
$$
{\cal A}_{CP} = \int  
{\frac
 {\left|{\cal M}\right|^2 - \left|{\cal \overline{M}}\right|^2}
 {\left|{\cal M}\right|^2 + \left|{\cal \overline{M}}\right|^2}
 \ dm^2_{ab}\ dm^2_{bc}  } 
 \ \left/ \int  dm^2_{ab}\ dm^2_{bc}\right. ,
$$
where ${\cal M}$ and ${\cal \overline{M}}$ are the $D^0$ and $\overline{D}^0$
Dalitz plot amplitudes.
This expression is less sensitive to $CP$ violation than the individual resonant
submodes reported in Ref.~\cite{cleoacp}.
\Table{cpv} reports the results for $CP$ violation.
No evidence of $CP$ violation has been observed.

\midtable{cpv}
\Caption
Dalitz-plot-integrated $CP$ violation.
\endCaption
\vglue -18pt
\centerline{
  \vbox{
    \halign{
  \centertab{#}&\centertab{#}&\centertab{#}\cr
  \tableheaddoublerule
  Experiment               & \hglue 0.1in Decay mode \hglue 0.3in       & ${\cal A}_{CP}$(\%)  \cr 
  \tableheadsinglerule
  CLEO\cite{Kopp:2000gv}   & $D^0 \to K^- \pi^+\pi^0$    & $-3.1\pm 8.6$ \cr
  CLEO\cite{cleob}         & $D^0 \to K^+ \pi^-\pi^0$    & $+9^{+22}_{-25}$ \cr
  CLEO\cite{cleoacp}       & $D^0 \to K^0_S \pi^+\pi^-$  & $-0.9\pm 2.1\!^{+1.0}_{-4.3}\!^{+1.3}_{-3.7}$ \cr
  CLEO\cite{cleopipipi0}   & $D^0 \to \pi^+\pi^-\pi^0$   & $+1^{+9}_{-7}\pm 9$ \cr
  \tableheaddoublerule
    }
  }
}
\endtable
The possibility of interference between $CP$--conserving and $CP$--violating
amplitudes provides a more sensitive probe of $CP$ violation. The constraints
on the square of the $CP$--violating amplitude obtained in the resonant submodes 
of $D^0 \to K^0_S \pi^+ \pi^-$ range form $(3.5$ to $28.4) \times 10^{-4}$ at 95\%
confidence level\cite{cleoacp}.

\medskip
\noindent{\boldface\bfit Guidance for Future Analyses:}
It is essential that a consistent formalism be adopted by all experiments that
report Charm Dalitz plot analyses. This is necessary for the Particle Data Group
to sensibly combine results from different experiments. 
Differences in the
parametrizations of the angular distribution $Z$, the barrier factors $B_L$ and 
the dynamical function $T_r$, as well as the set of 
resonances $r$, complicate the comparison of results from 
different experiments. 

As the sizes of the data samples increase, the dynamical model which describes
the intermediate resonances must be improved.
Replacing the sum of Breit-Wigner amplitudes with a unitary parametrization
such as the $K$-matrix is important. Of course, other dynamical models should
be studied. In particular, the phase shift associated with $\pi^+\pi^+$ in $D^+_{(s)} \to K^-\pi^+\pi^+, \pi^-\pi^+\pi^+$ should not be neglected and momentum-dependent form-factors in \Eq{vec} should be considered.


Finally, for a consistent picture of the $\pi\pi$ and $K\pi$ scalar mesons to
emerge from charm Dalitz plot analyses many decays must be evaluated simultaneously.
This will allow resonances decaying to different final states to be analyzed
together. The same set of intermediate states can be used for all final
states and input (masses, widths, couplings, spectra) from 
scattering experiments like $\pi^- p \to \pi^0\pi^0 n$, radiative decays such as $\phi, J/\psi \to \gamma \pi\pi$ and semileptonic decays such as $D \to K\pi\ell\nu$ can be consistently incorporated.

\minirefs
\ListReferences
\endmini

\vfill\eject
\end

%% file: minimacros.tex
%

%
%
%

\def\tableheaddoublerule{\noalign{\vglue2pt\hrule\vskip3pt\hrule\smallskip}}
\def\tableheadsinglerule{\noalign{\medskip\hrule\smallskip}}
\def\centertab#1{\hfil#1\hfil}

\def\lefttab#1{#1\hfil}

\def\mbox#1{{\ifmmode#1\else$#1$\fi}}

%
%
\newskip\doublecolskip 
\def\BeginMiniGeneral%
{%
    \bgroup
%
%
    \par\raggedright\raggedbottom\twelvepoint%
    \parindent = 20pt \leftskip = 0pt \rightskip = 0pt%
%
%
    \doublecolskip=.3333em plus .3333em minus .1em%
    \spaceskip=\doublecolskip%
    \baselineskip 17pt%
}

\def\endmini{\par\refReset\eqReset%
    \global\tabnum=0\global\fignum=0\Flush\egroup} 
\def\endminiinsert{\par\baselineskip 17pt\Flush\egroup}
%
%
\def\miniinsert#1{%
  \ifnum\MiniFinal=0
     {\def\Flush{}
     \input #1
     }
  \else
     \input #1
  \fi
}
\ifnum\MiniFinal = 0
%
\ATunlock       
%
\def\refFormat{\catcode`\^^M=10}

\superrefsfalse 

\def\Table#1{Table~\use{Tb.#1}}


%
%
\newcount\footnum        \footnum=0
\def\eqReset{\global\eqnum=0}
%
\def\bumpupequationnumber{\global\advance\eqnum by 1\relax}
\def\bumpupchapternumber{\global\advance\chapternum by 1\relax}
\def\bumpupsectionnumber{\global\advance\sectionnum by 1\relax}
\def\bumpupsubsectionnumber{\global\advance\subsectionnum by 1\relax}
\def\bumpupfigurenumber{\global\advance\fignum by 1\relax}
\def\bumpuptablenumber{\global\advance\tabnum by 1\relax}
\def\bumpdownequationnumber{\global\advance\eqnum by -1\relax}
\def\bumpdownchapternumber{\global\advance\chapternum by -1\relax}
\def\bumpdownsectionnumber{\global\advance\sectionnum by -1\relax}
\def\bumpdownsubsectionnumber{\global\advance\subsectionnum by -1\relax}
\def\bumpdownfigurenumber{\global\advance\fignum by -1\relax}
\def\bumpdowntablenumber{\global\advance\tabnum by -1\relax}
\def\labelfigure#1{\tag{Fg.#1}{\the\chapternum.\the\fignum}}
\def\labeltable#1{\tag{Tb.#1}{\the\chapternum.\the\tabnum}}
\def\labelequation#1{\tag{Eq.#1}{\the\chapternum.\the\eqnum}}

\def\labelsection#1{\tag{Sec.#1}{\the\chapternum.\the\sectionnum}}
\def\labelsubsection#1{\tag{Sec.#1}{\the\chapternum.\the\sectionnum.%
\the\subsectionnum}}
\def\labelsubsubsection#1{\tag{Sec.#1}{\the\chapternum.\the\sectionnum.%
\the\subsectionnum.\the\subsubsectionnum}}
%
%

%
\newskip\lefteqnside
\newskip\righteqnside
\newdimen\lefteqnsidedimen \lefteqnsidedimen=22pt 
\lefteqnside =0pt\relax\righteqnside=0pt plus 1fil  
\lefteqnside=0pt plus 1fil\relax\righteqnside =0pt 
\lefteqnside =0pt plus 1fil 
\righteqnside=0pt plus 1fil 
\lefteqnsidedimen=30pt 
\lefteqnside =30pt 
\righteqnside=0pt plus 1fil  
\def\RPPdisplaylines#1{
      \@EQNcr                             
    \openup 2\jot
    \displ@y                            
   \halign{\hbox to \displaywidth{$\relax\hskip\lefteqnside{\displaystyle##}%
               \hskip\righteqnside$}%
   &\llap{$\relax\@@EQN{##}$}\crcr      
    #1\crcr}
    \@EQNuncr                          
    }


\long\def\RPPalign#1{
  \@EQNcr                               
    \openup 2\jot
   \displ@y                              
     \tabskip=\lefteqnside                 
   \halign to\displaywidth{
   \hfil$\relax\displaystyle{##}$
     \tabskip=0pt                        
   &$\leavevmode\relax\displaystyle{{}##}$\hfil     
     \tabskip=\righteqnside                 
  &\llap{$\relax\@@EQN{##}$}
     \tabskip=0pt\crcr                   
    #1\crcr}
   }


\def\RPPdoublealign#1{
   \@EQNcr                              
    \openup 2\jot
   \displ@y                             
     \tabskip=\lefteqnside                 
   \halign to\displaywidth{
      \hfil$\relax\displaystyle{##}$
      \tabskip=0pt                      
   &$\relax\displaystyle{{}##}$\hfil
      \tabskip=0pt                      
   &$\relax\displaystyle{{}##}$\hfil
     \tabskip=\righteqnside                 
   &\llap{$\relax\@@EQN{##}$}
      \tabskip=0pt\crcr                 
   #1\crcr}
   \@EQNuncr                          
   }%


\def\today{\ifcase\month\or
  January\or February\or March\or April\or May\or June\or
  July\or August\or September\or October\or November\or December\fi
  \space\number\day, \number\year}
\def\fildec#1{\ifnum#1<10 0\fi\the#1}
\newcount\hour \newcount\minute
\def\TimeOfDay%
{
   \hour\time\divide\hour by 60
   \minute-\hour\multiply\minute by 60 \advance\minute\time
   \fildec\hour:\fildec\minute
}

\ATlock         

%
%
\def\elevenfonts{%
   \global\font\elevenrm=cmr10 scaled \magstephalf
   \global\font\eleveni=cmmi10 scaled \magstephalf
   \global\font\elevensy=cmsy10 scaled \magstephalf
   \global\font\elevenex=cmex10 scaled \magstephalf
   \global\font\elevenbf=cmbx10 scaled \magstephalf
   \global\font\elevensl=cmsl10 scaled \magstephalf
   \global\font\eleventt=cmtt10 scaled \magstephalf
   \global\font\elevenit=cmti10 scaled \magstephalf
   \global\font\elevenss=cmss10 scaled \magstephalf
   \global\font\elevenbxti=cmbxti10 scaled \magstephalf
   \skewchar\eleveni='177
   \skewchar\elevensy='60
   \hyphenchar\eleventt=-1
   \moreelevenfonts                            
   \gdef\elevenfonts{\relax}}%

\def\moreelevenfonts{\relax}                    

%
%
\def\twelvefonts{
   \global\font\twelverm=cmr12 
   \global\font\twelvei=cmmi10 scaled \magstep1
   \global\font\twelvesy=cmsy10 scaled \magstep1
   \global\font\twelveex=cmex10 scaled \magstep1
   \global\font\twelvebf=cmbx12
   \global\font\twelvesl=cmsl12
   \global\font\twelvett=cmtt12
   \global\font\twelveit=cmti12
   \global\font\twelvess=cmss12
   \skewchar\twelvei='177
   \skewchar\twelvesy='60
   \hyphenchar\twelvett=-1
   \moretwelvefonts                             
   \gdef\twelvefonts{\relax}}

\def\moretwelvefonts{\relax}

%
%
\newskip\strutskip
\def\strut{\vrule height 0.8\strutskip depth 0.3\strutskip width 0pt}

\message{10pt,}
\def\tenpoint{
   \def\rm{\fam0\tenrm}%
   \textfont0=\tenrm\scriptfont0=\eightrm\scriptscriptfont0=\sevenrm
   \textfont1=\teni\scriptfont1=\eighti\scriptscriptfont1=\seveni
   \textfont2=\tensy\scriptfont2=\eightsy\scriptscriptfont2=\sevensy
%
%
   \textfont3=\tenex\scriptfont3=\eightex\scriptscriptfont3=\sevenex
   \textfont4=\tenit\scriptfont4=\eightit\scriptscriptfont4=\sevenit
   \textfont\itfam=\tenit\def\it{\fam\itfam\tenit}%
   \textfont\slfam=\tensl\def\sl{\fam\slfam\tensl}%
   \textfont\ttfam=\tentt\def\tt{\fam\ttfam\tentt}%
   \textfont\bffam=\tenbf
   \scriptfont\bffam=\eightbf
   \scriptscriptfont\bffam=\sevenbf\def\bf{\fam\bffam\tenbf}%
%
%
   \def\mib{%
      \tenmibfonts
      \textfont0=\tenbf\scriptfont0=\eightbf
      \scriptscriptfont0=\sevenbf
      \textfont1=\tenmib\scriptfont1=\eighti
      \scriptscriptfont1=\seveni
      \textfont2=\tenbsy\scriptfont2=\eightsy
      \scriptscriptfont2=\sevensy}%
   \def\scr{\scrfonts
      \global\textfont\scrfam=\tenscr\fam\scrfam\tenscr}%
   \tt\ttglue=.5emplus.25emminus.15em
   \normalbaselineskip=12pt
   \setbox\strutbox=\hbox{\vrule height 8.5pt depth 3.5pt width 0pt}%
   \normalbaselines\rm\singlespaced
   \let\emphfont=\it
   \let\bfit=\tenbxti
   \let\itbf=\tenbxti
   \let\boldface=\boldtenpoint
\def\setstrut{\strutskip = \baselineskip}\setstrut%
\def\strut{\vrule height 0.7\strutskip depth 0.3\strutskip width 0pt}%
      }%

%
%
\message{11pt,}
\def\elevenpoint{\elevenfonts           
   \def\rm{\fam0\elevenrm}%
   \textfont0=\elevenrm\scriptfont0=\eightrm\scriptscriptfont0=\sevenrm
   \textfont1=\eleveni\scriptfont1=\eighti\scriptscriptfont1=\seveni
   \textfont2=\elevensy\scriptfont2=\eightsy\scriptscriptfont2=\sevensy
   \textfont3=\elevenex\scriptfont3=\elevenex\scriptscriptfont3=\elevenex
   \textfont\itfam=\elevenit\def\it{\fam\itfam\elevenit}%
   \textfont\slfam=\elevensl\def\sl{\fam\slfam\elevensl}%
   \textfont\ttfam=\eleventt\def\tt{\fam\ttfam\eleventt}%
   \textfont\bffam=\elevenbf
   \scriptfont\bffam=\eightbf
   \scriptscriptfont\bffam=\sevenbf\def\bf{\fam\bffam\elevenbf}%
   \def\mib{%
      \elevenmibfonts
      \textfont0=\elevenbf\scriptfont0=\eightbf
      \scriptscriptfont0=\sevenbf
      \textfont1=\elevenmib\scriptfont1=\eightmib
      \scriptscriptfont1=\sevenmib
      \textfont2=\elevenbsy\scriptfont2=\eightsy
      \scriptscriptfont2=\sevensy}%
   \def\scr{\scrfonts
      \global\textfont\scrfam=\elevenscr\fam\scrfam\elevenscr}%
   \tt\ttglue=.5emplus.25emminus.15em
   \normalbaselineskip=13pt
   \setbox\strutbox=\hbox{\vrule height 9pt depth 4pt width 0pt}%
   \let\emphfont=\it
   \let\bfit=\elevenbxti
   \let\itbf=\elevenbxti
\def\setstrut{\strutskip = \baselineskip}\setstrut%
\def\strut{\vrule height 0.7\strutskip depth 0.3\strutskip width 0pt}%
   \let\boldface=\boldelevenpoint
   \normalbaselines\rm\singlespaced}%

\message{12pt,}
\def\twelvepoint{\twelvefonts\ninefonts 
   \def\rm{\fam0\twelverm}%
   \textfont0=\twelverm\scriptfont0=\ninerm\scriptscriptfont0=\sevenrm
   \textfont1=\twelvei\scriptfont1=\ninei\scriptscriptfont1=\seveni
   \textfont2=\twelvesy\scriptfont2=\ninesy\scriptscriptfont2=\sevensy
   \textfont3=\twelveex\scriptfont3=\twelveex\scriptscriptfont3=\twelveex
   \textfont\itfam=\twelveit\def\it{\fam\itfam\twelveit}%
   \textfont\slfam=\twelvesl\def\sl{\fam\slfam\twelvesl}%
   \textfont\ttfam=\twelvett\def\tt{\fam\ttfam\twelvett}%
   \textfont\bffam=\twelvebf
   \scriptfont\bffam=\ninebf
   \scriptscriptfont\bffam=\sevenbf\def\bf{\fam\bffam\twelvebf}%
   \def\mib{%
      \twelvemibfonts\tenmibfonts
      \textfont0=\twelvebf\scriptfont0=\ninebf
      \scriptscriptfont0=\sevenbf
      \textfont1=\twelvemib\scriptfont1=\ninemib
      \scriptscriptfont1=\sevenmib
      \textfont2=\twelvebsy\scriptfont2=\ninesy
      \scriptscriptfont2=\sevensy}%
   \def\scr{\scrfonts
      \global\textfont\scrfam=\twelvescr\fam\scrfam\twelvescr}%
   \tt\ttglue=.5emplus.25emminus.15em
   \normalbaselineskip=14pt
   \setbox\strutbox=\hbox{\vrule height 10pt depth 4pt width 0pt}%
   \let\emphfont=\it
   \let\bfit=\twelvebxti
   \let\itbf=\twelvebxti
\def\setstrut{\strutskip = \baselineskip}\setstrut%
\def\strut{\vrule height 0.7\strutskip depth 0.3\strutskip width 0pt}%
   \let\boldface=\boldtwelvepoint
   \normalbaselines\rm\singlespaced}%

%
	\font\sevenex=cmex10 scaled 667
	\font\eightex=cmex10 scaled 800

	\font\eightbf cmbx8
	\font\eighti cmmi8
	\font\eightit cmti8
	\font\sevenit cmti7
	\font\eightrm cmr8
	\font\eightsy cmsy8
   \skewchar\eightsy='60
	\font\eightmib cmmib8
   \skewchar\eightmib='177
	\font\sevenmib cmmib7
   \skewchar\sevenmib='177
	\font\ninemib cmmib9
   \skewchar\ninemib='177
	\font\tenbxti cmbxti10
	\font\twelvebxti cmbxti10 scaled 1200

\def\extrasportsfonts{%
	\font\eightbsy cmbsy10 scaled 800
	\font\eightssbf cmssbx10 scaled 800
	\font\eightssi cmssi8
	\font\niness cmss9
	\font\msxmten=msam10 
	\font\tenssbf cmssbx10
	\font\elevenssbf cmssbx10 scaled 1095
	\font\twelvessbf cmssbx10 scaled 1200
\font\Bigbf cmbx12 scaled 1440
\font\Bigti cmti12 scaled \magstep2
\let\boldhelv \tenssbf
\let\helv \tenss
\def\interheadbf{\tenbf\bf\mib}
\let\hf\boldhead
\let\headfont\boldhead
\gdef\extrasportsfonts{\relax}}%
\def\extraminifonts{%
   \font\msxmtwelve=msam10  scaled 1200 
\gdef\extraminifonts{\relax}}%
%

\gdef\Tbf{\twelvepoint\bf\mib}
\gdef\tbf{\tenpoint\bf\mib}
%
%
\def\boldtenpoint{\tenpoint\bf\mib%
   \textfont\itfam=\tenbxti\def\it{\fam\itfam\tenbxti}%
\relax}
\def\boldelevenpoint{\elevenpoint\bf\mib%
   \textfont\itfam=\elevenbxti\def\it{\fam\itfam\elevenbxti}%
\relax}
\def\boldtwelvepoint{\twelvepoint\bf\mib%
   \textfont\itfam=\twelvebxti\def\it{\fam\itfam\twelvebxti}%
\relax}
\def\etal{\hbox{\it et~al.}}
\tenpoint\rm

     
\def\enumalpha{
   \def\setenumlead{\def\enumlead{}}
   \def\enumcur{\ifcase\enumDepth               
      \or\letterN{\the\enumcnt}
      \or{\XA\romannumeral\number\enumcnt}
      \or{\XA\number\enumcnt}
      \else $\bullet$\space\fi}
   }

     
\def\enumroman{
   \def\setenumlead{\def\enumlead{}}
   \def\enumcur{\ifcase\enumDepth               
      \or{\XA\romannumeral\number\enumcnt}
      \or\letterN{\the\enumcnt}
      \or{\XA\number\enumcnt}
      \else $\bullet$\space\fi}
   }
\def\anp#1,#2(#3){{\rm Adv.\ Nucl.\ Phys.\ }{\bf #1}, {\rm#2} {\rm(#3)}}
\def\aip#1,#2(#3){{\rm Am.\ Inst.\ Phys.\ }{\bf #1}, {\rm#2} {\rm(#3)}}
\def\aj#1,#2(#3){{\rm Astrophys.\ J.\ }{\bf #1}, {\rm#2} {\rm(#3)}}
\def\ajs#1,#2(#3){{\rm Astrophys.\ J.\ Supp.\ }{\bf #1}, {\rm#2} {\rm(#3)}}
\def\ajl#1,#2(#3){{\rm Astrophys.\ J.\ Lett.\ }{\bf #1}, {\rm#2} {\rm(#3)}}
\def\ajp#1,#2(#3){{\rm Am.\ J.\ Phys.\ }{\bf #1}, {\rm#2} {\rm(#3)}}
\def\apny#1,#2(#3){{\rm Ann.\ Phys.\ (NY)\ }{\bf #1}, {\rm#2} {\rm(#3)}}
\def\apnyB#1,#2(#3){{\rm Ann.\ Phys.\ (NY)\ }{\bf B#1}, {\rm#2} {\rm(#3)}}
\def\apD#1,#2(#3){{\rm Ann.\ Phys.\ }{\bf D#1}, {\rm#2} {\rm(#3)}}
\def\ap#1,#2(#3){{\rm Ann.\ Phys.\ }{\bf #1}, {\rm#2} {\rm(#3)}}
\def\ass#1,#2(#3){{\rm Ap.\ Space Sci.\ }{\bf #1}, {\rm#2} {\rm(#3)}}
\def\astropp#1,#2(#3)%
    {{\rm Astropart.\ Phys.\ }{\bf #1}, {\rm#2} {\rm(#3)}}
\def\aap#1,#2(#3)%
    {{\rm Astron.\ \& Astrophys.\ }{\bf #1}, {\rm#2} {\rm(#3)}}
\def\araa#1,#2(#3)%
    {{\rm Ann.\ Rev.\ Astron.\ Astrophys.\ }{\bf #1}, {\rm#2} {\rm(#3)}}
\def\arnps#1,#2(#3)%
    {{\rm Ann.\ Rev.\ Nucl.\ and Part.\ Sci.\ }{\bf #1}, {\rm#2} {\rm(#3)}}
\def\arns#1,#2(#3)%
   {{\rm Ann.\ Rev.\ Nucl.\ Sci.\ }{\bf #1}, {\rm#2} {\rm(#3)}}
\def\cpc#1,#2(#3){{\rm Comp.\ Phys.\ Comm.\ }{\bf #1}, {\rm#2} {\rm(#3)}}
\def\cjp#1,#2(#3){{\rm Can.\ J.\ Phys.\ }{\bf #1}, {\rm#2} {\rm(#3)}}
\def\cmp#1,#2(#3){{\rm Commun.\ Math.\ Phys.\ }{\bf #1}, {\rm#2} {\rm(#3)}}
\def\cnpp#1,#2(#3)%
   {{\rm Comm.\ Nucl.\ Part.\ Phys.\ }{\bf #1}, {\rm#2} {\rm(#3)}}
\def\cnppA#1,#2(#3)%
   {{\rm Comm.\ Nucl.\ Part.\ Phys.\ }{\bf A#1}, {\rm#2} {\rm(#3)}}
\def\epjC#1,#2(#3){{\rm Eur.\ Phys.\ J.\ }{\bf C#1}, {\rm#2} {\rm(#3)}}
\def\el#1,#2(#3){{\rm Europhys.\ Lett.\ }{\bf #1}, {\rm#2} {\rm(#3)}}
\def\hpa#1,#2(#3){{\rm Helv.\ Phys.\ Acta }{\bf #1}, {\rm#2} {\rm(#3)}}
\def\ieeetNS#1,#2(#3)%
    {{\rm IEEE Trans.\ }{\bf NS#1}, {\rm#2} {\rm(#3)}}
\def\IEEE #1,#2(#3)%
    {{\rm IEEE }{\bf #1}, {\rm#2} {\rm(#3)}}
\def\ijar#1,#2(#3)%
  {{\rm Int.\ J.\ of Applied Rad.\ } {\bf #1}, {\rm#2} {\rm(#3)}}
\def\ijari#1,#2(#3)%
  {{\rm Int.\ J.\ of Applied Rad.\ and Isotopes\ } {\bf #1}, {\rm#2} {\rm(#3)}}
\def\jcp#1,#2(#3){{\rm J.\ Chem.\ Phys.\ }{\bf #1}, {\rm#2} {\rm(#3)}}
\def\jgr#1,#2(#3){{\rm J.\ Geophys.\ Res.\ }{\bf #1}, {\rm#2} {\rm(#3)}}
\def\jetp#1,#2(#3){{\rm Sov.\ Phys.\ JETP\ }{\bf #1}, {\rm#2} {\rm(#3)}}
\def\jetpl#1,#2(#3)%
   {{\rm Sov.\ Phys.\ JETP Lett.\ }{\bf #1}, {\rm#2} {\rm(#3)}}
\def\jpA#1,#2(#3){{\rm J.\ Phys.\ }{\bf A#1}, {\rm#2} {\rm(#3)}}
\def\jpG#1,#2(#3){{\rm J.\ Phys.\ }{\bf G#1}, {\rm#2} {\rm(#3)}}
\def\jpamg#1,#2(#3)%
    {{\rm J.\ Phys.\ A: Math.\ and Gen.\ }{\bf #1}, {\rm#2} {\rm(#3)}}
\def\jpcrd#1,#2(#3)%
    {{\rm J.\ Phys.\ Chem.\ Ref.\ Data\ } {\bf #1}, {\rm#2} {\rm(#3)}}
\def\jpsj#1,#2(#3){{\rm J.\ Phys.\ Soc.\ Jpn.\ }{\bf G#1}, {\rm#2} {\rm(#3)}}
\def\lnc#1,#2(#3){{\rm Lett.\ Nuovo Cimento\ } {\bf #1}, {\rm#2} {\rm(#3)}}
\def\nature#1,#2(#3){{\rm Nature} {\bf #1}, {\rm#2} {\rm(#3)}}
\def\nc#1,#2(#3){{\rm Nuovo Cimento} {\bf #1}, {\rm#2} {\rm(#3)}}
\def\nim#1,#2(#3)%
   {{\rm Nucl.\ Instrum.\ Methods\ }{\bf #1}, {\rm#2} {\rm(#3)}}
\def\nimA#1,#2(#3)%
    {{\rm Nucl.\ Instrum.\ Methods\ }{\bf A#1}, {\rm#2} {\rm(#3)}}
\def\nimB#1,#2(#3)%
    {{\rm Nucl.\ Instrum.\ Methods\ }{\bf B#1}, {\rm#2} {\rm(#3)}}
\def\np#1,#2(#3){{\rm Nucl.\ Phys.\ }{\bf #1}, {\rm#2} {\rm(#3)}}
\def\mnras#1,#2(#3){{\rm ASK.\ GEORGE.\ }{\bf #1}, {\rm#2} {\rm(#3)}}
\def\medp#1,#2(#3){{\rm Med.\ Phys.\ }{\bf #1}, {\rm#2} {\rm(#3)}}
\def\mplA#1,#2(#3){{\rm Mod.\ Phys.\ Lett.\ }{\bf A#1}, {\rm#2} {\rm(#3)}}
\def\npA#1,#2(#3){{\rm Nucl.\ Phys.\ }{\bf A#1}, {\rm#2} {\rm(#3)}}
\def\npB#1,#2(#3){{\rm Nucl.\ Phys.\ }{\bf B#1}, {\rm#2} {\rm(#3)}}
\def\npBps#1,#2(#3){{\rm Nucl.\ Phys.\ (Proc.\ Supp.) }{\bf B#1},
{\rm#2} {\rm(#3)}}
\def\pasp#1,#2(#3){{\rm Pub.\ Astron.\ Soc.\ Pac.\ }{\bf #1}, {\rm#2} {\rm(#3)}}
\def\pl#1,#2(#3){{\rm Phys.\ Lett.\ }{\bf #1}, {\rm#2} {\rm(#3)}}
\def\fp#1,#2(#3){{\rm Fortsch.\ Phys.\ }{\bf #1}, {\rm#2} {\rm(#3)}}
\def\ijmpA#1,#2(#3)%
   {{\rm Int.\ J.\ Mod.\ Phys.\ }{\bf A#1}, {\rm#2} {\rm(#3)}}
\def\ijmpE#1,#2(#3)%
   {{\rm Int.\ J.\ Mod.\ Phys.\ }{\bf E#1}, {\rm#2} {\rm(#3)}}
\def\plB#1,#2(#3){{\rm Phys.\ Lett.\ }{\bf B#1}, {\rm#2} {\rm(#3)}}
\def\pnasus#1,#2(#3)%
   {{\it Proc.\ Natl.\ Acad.\ Sci.\ \rm (US)}{B#1}, {\rm#2} {\rm(#3)}}
\def\ppsA#1,#2(#3){{\rm Proc.\ Phys.\ Soc.\ }{\bf A#1}, {\rm#2} {\rm(#3)}}
\def\ppsB#1,#2(#3){{\rm Proc.\ Phys.\ Soc.\ }{\bf B#1}, {\rm#2} {\rm(#3)}}
\def\pr#1,#2(#3){{\rm Phys.\ Rev.\ }{\bf #1}, {\rm#2} {\rm(#3)}}
\def\prA#1,#2(#3){{\rm Phys.\ Rev.\ }{\bf A#1}, {\rm#2} {\rm(#3)}}
\def\prB#1,#2(#3){{\rm Phys.\ Rev.\ }{\bf B#1}, {\rm#2} {\rm(#3)}}
\def\prC#1,#2(#3){{\rm Phys.\ Rev.\ }{\bf C#1}, {\rm#2} {\rm(#3)}}
\def\prD#1,#2(#3){{\rm Phys.\ Rev.\ }{\bf D#1}, {\rm#2} {\rm(#3)}}
\def\prept#1,#2(#3){{\rm Phys.\ Reports\ } {\bf #1}, {\rm#2} {\rm(#3)}}
\def\preptC#1,#2(#3){{\rm Phys.\ Reports\ } {\bf C#1}, {\rm#2} {\rm(#3)}}
\def\prslA#1,#2(#3)%
   {{\rm Proc.\ Royal Soc.\ London }{\bf A#1}, {\rm#2} {\rm(#3)}}
\def\prl#1,#2(#3){{\rm Phys.\ Rev.\ Lett.\ }{\bf #1}, {\rm#2} {\rm(#3)}}
\def\ps#1,#2(#3){{\rm Phys.\ Scripta\ }{\bf #1}, {\rm#2} {\rm(#3)}}
\def\ptp#1,#2(#3){{\rm Prog.\ Theor.\ Phys.\ }{\bf #1}, {\rm#2} {\rm(#3)}}
\def\ppnp#1,#2(#3)%
	{{\rm Prog.\ in Part.\ Nucl.\ Phys.\ }{\bf #1}, {\rm#2} {\rm(#3)}}
\def\ptps#1,#2(#3)%
   {{\rm Prog.\ Theor.\ Phys.\ Supp.\ }{\bf #1}, {\rm#2} {\rm(#3)}}
\def\pw#1,#2(#3){{\rm Part.\ World\ }{\bf #1}, {\rm#2} {\rm(#3)}}
\def\pzetf#1,#2(#3)%
   {{\rm Pisma Zh.\ Eksp.\ Teor.\ Fiz.\ }{\bf #1}, {\rm#2} {\rm(#3)}}
\def\rgss#1,#2(#3){{\rm Revs.\ Geophysics \& Space Sci.\ }{\bf #1},
        {\rm#2} {\rm(#3)}}
\def\rmp#1,#2(#3){{\rm Rev.\ Mod.\ Phys.\ }{\bf #1}, {\rm#2} {\rm(#3)}}
\def\rnc#1,#2(#3){{\rm Riv.\ Nuovo Cimento\ } {\bf #1}, {\rm#2} {\rm(#3)}}
\def\rpp#1,#2(#3)%
    {{\rm Rept.\ on Prog.\ in Phys.\ }{\bf #1}, {\rm#2} {\rm(#3)}}
\def\science#1,#2(#3){{\rm Science\ } {\bf #1}, {\rm#2} {\rm(#3)}}
\def\sjnp#1,#2(#3)%
   {{\rm Sov.\ J.\ Nucl.\ Phys.\ }{\bf #1}, {\rm#2} {\rm(#3)}}
\def\sjpn#1,#2(#3)%
   {{\rm Sov.\ J.\ Part.\ Nucl.\ }{\bf #1}, {\rm#2} {\rm(#3)}}
\def\panp#1,#2(#3)%
   {{\rm Phys.\ Atom.\ Nucl.\ }{\bf #1}, {\rm#2} {\rm(#3)}}
\def\spu#1,#2(#3){{\rm Sov.\ Phys.\ Usp.\ }{\bf #1}, {\rm#2} {\rm(#3)}}
\def\surveyHEP#1,#2(#3)%
    {{\rm Surv.\ High Energy Physics\ } {\bf #1}, {\rm#2} {\rm(#3)}}
\def\yf#1,#2(#3){{\rm Yad.\ Fiz.\ }{\bf #1}, {\rm#2} {\rm(#3)}}
\def\zetf#1,#2(#3)%
   {{\rm Zh.\ Eksp.\ Teor.\ Fiz.\ }{\bf #1}, {\rm#2} {\rm(#3)}}
\def\zp#1,#2(#3){{\rm Z.~Phys.\ }{\bf #1}, {\rm#2} {\rm(#3)}}
\def\zpA#1,#2(#3){{\rm Z.~Phys.\ }{\bf A#1}, {\rm#2} {\rm(#3)}}
\def\zpC#1,#2(#3){{\rm Z.~Phys.\ }{\bf C#1}, {\rm#2} {\rm(#3)}}
%
%

%

%

%

%

%

%

%

%

%

%

%

%

%

%

%

%

%

%

    \hsize 4.5 truein
    \vsize 9 truein%
    \def\Flush{\vfill\eject}
    \def\beginmini{\BeginMiniGeneral}
    \def\minihead#1%
    {%
        \global\headline={\bf #1\quad\the\pageno\hss\today\quad\TimeOfDay}%
        \global\footline={\bf #1\quad\the\pageno\hss}%
    }
    \def\miniheadinsert#1{} 
\def\PsfigVersion{1.9}
\ifx\undefined\psfig\else \fi

%

\let\LaTeXAtSign=\@
\let\@=\relax
\edef\psfigRestoreAt{\catcode`\@=\number\catcode`@\relax}
\catcode`\@=11\relax
\newwrite\@unused
\def\ps@typeout#1{{\let\protect\string\immediate\write\@unused{#1}}}
\ps@typeout{psfig/tex \PsfigVersion}


\def\figurepath{./}

%
%
\def\@nnil{\@nil}
\def\@empty{}
\def\@psdonoop#1\@@#2#3{}
\def\@psdo#1:=#2\do#3{\edef\@psdotmp{#2}\ifx\@psdotmp\@empty \else
    \expandafter\@psdoloop#2,\@nil,\@nil\@@#1{#3}\fi}
\def\@psdoloop#1,#2,#3\@@#4#5{\def#4{#1}\ifx #4\@nnil \else
       #5\def#4{#2}\ifx #4\@nnil \else#5\@ipsdoloop #3\@@#4{#5}\fi\fi}
\def\@ipsdoloop#1,#2\@@#3#4{\def#3{#1}\ifx #3\@nnil 
       \let\@nextwhile=\@psdonoop \else
      #4\relax\let\@nextwhile=\@ipsdoloop\fi\@nextwhile#2\@@#3{#4}}
\def\@tpsdo#1:=#2\do#3{\xdef\@psdotmp{#2}\ifx\@psdotmp\@empty \else
    \@tpsdoloop#2\@nil\@nil\@@#1{#3}\fi}
\def\@tpsdoloop#1#2\@@#3#4{\def#3{#1}\ifx #3\@nnil 
       \let\@nextwhile=\@psdonoop \else
      #4\relax\let\@nextwhile=\@tpsdoloop\fi\@nextwhile#2\@@#3{#4}}
%
\ifx\undefined\fbox
\newdimen\fboxrule
\newdimen\fboxsep
\newdimen\ps@tempdima
\newbox\ps@tempboxa
\fboxsep = 3pt
\fboxrule = .4pt
\long\def\fbox#1{\leavevmode\setbox\ps@tempboxa\hbox{#1}\ps@tempdima\fboxrule
    \advance\ps@tempdima \fboxsep \advance\ps@tempdima \dp\ps@tempboxa
   \hbox{\lower \ps@tempdima\hbox
  {\vbox{\hrule height \fboxrule
          \hbox{\vrule width \fboxrule \hskip\fboxsep
          \vbox{\vskip\fboxsep \box\ps@tempboxa\vskip\fboxsep}\hskip 
                 \fboxsep\vrule width \fboxrule}
                 \hrule height \fboxrule}}}}
\fi
%
%
\newread\ps@stream
\newif\ifnot@eof       
\newif\if@noisy        
\newif\if@atend        
\newif\if@psfile       
%
%
{\catcode`\%=12\global\gdef\epsf@start{
\def\epsf@PS{PS}
\def\epsf@getbb#1{%
%
%
\openin\ps@stream=#1
\ifeof\ps@stream\ps@typeout{Error, File #1 not found}\else
%
%
   {\not@eoftrue \chardef\other=12
    \def\do##1{\catcode`##1=\other}\dospecials \catcode`\ =10
    \loop
       \if@psfile
	  \read\ps@stream to \epsf@fileline
       \else{
	  \obeyspaces
          \read\ps@stream to \epsf@tmp\global\let\epsf@fileline\epsf@tmp}
       \fi
       \ifeof\ps@stream\not@eoffalse\else
%
%
       \if@psfile\else
       \expandafter\epsf@test\epsf@fileline:. \\%
       \fi
%
%
          \expandafter\epsf@aux\epsf@fileline:. \\%
       \fi
   \ifnot@eof\repeat
   }\closein\ps@stream\fi}%
%
%
\long\def\epsf@test#1#2#3:#4\\{\def\epsf@testit{#1#2}
			\ifx\epsf@testit\epsf@start\else
\ps@typeout{Warning! File does not start with `\epsf@start'.  It may not be a PostScript file.}
			\fi
			\@psfiletrue} 
%
%
{\catcode`\%=12\global\let\epsf@percent=
%
%
%
\long\def\epsf@aux#1#2:#3\\{\ifx#1\epsf@percent
   \def\epsf@testit{#2}\ifx\epsf@testit\epsf@bblit
	\@atendfalse
        \epsf@atend #3 . \\%
	\if@atend	
	   \if@verbose{
		\ps@typeout{psfig: found `(atend)'; continuing search}
	   }\fi
        \else
        \epsf@grab #3 . . . \\%
        \not@eoffalse
        \global\no@bbfalse
        \fi
   \fi\fi}%
%
%
\def\epsf@grab #1 #2 #3 #4 #5\\{%
   \global\def\epsf@llx{#1}\ifx\epsf@llx\empty
      \epsf@grab #2 #3 #4 #5 .\\\else
   \global\def\epsf@lly{#2}%
   \global\def\epsf@urx{#3}\global\def\epsf@ury{#4}\fi}%
%
%
\def\epsf@atendlit{(atend)} 
\def\epsf@atend #1 #2 #3\\{%
   \def\epsf@tmp{#1}\ifx\epsf@tmp\empty
      \epsf@atend #2 #3 .\\\else
   \ifx\epsf@tmp\epsf@atendlit\@atendtrue\fi\fi}


\chardef\psletter = 11 
\chardef\other = 12

\newif \ifdebug 
\newif\ifc@mpute 
\c@mputetrue 

\let\then = \relax
\def\r@dian{pt }
\let\r@dians = \r@dian
\let\dimensionless@nit = \r@dian
\let\dimensionless@nits = \dimensionless@nit
\def\internal@nit{sp }
\let\internal@nits = \internal@nit
\newif\ifstillc@nverging
\def \Mess@ge #1{\ifdebug \then \message {#1} \fi}

{ 
	\catcode `\@ = \psletter
	\gdef \nodimen {\expandafter \n@dimen \the \dimen}
	\gdef \term #1 #2 #3%
	       {\edef \t@ {\the #1}
		\edef \t@@ {\expandafter \n@dimen \the #2\r@dian}%
		\t@rm {\t@} {\t@@} {#3}%
	       }
	\gdef \t@rm #1 #2 #3%
	       {{%
		\count 0 = 0
		\dimen 0 = 1 \dimensionless@nit
		\dimen 2 = #2\relax
		\Mess@ge {Calculating term #1 of \nodimen 2}%
		\loop
		\ifnum	\count 0 < #1
		\then	\advance \count 0 by 1
			\Mess@ge {Iteration \the \count 0 \space}%
			\Multiply \dimen 0 by {\dimen 2}%
			\Mess@ge {After multiplication, term = \nodimen 0}%
			\Divide \dimen 0 by {\count 0}%
			\Mess@ge {After division, term = \nodimen 0}%
		\repeat
		\Mess@ge {Final value for term #1 of 
				\nodimen 2 \space is \nodimen 0}%
		\xdef \Term {#3 = \nodimen 0 \r@dians}%
		\aftergroup \Term
	       }}
	\catcode `\p = \other
	\catcode `\t = \other
	\gdef \n@dimen #1pt{#1} 
}

\def \Divide #1by #2{\divide #1 by #2} 

\def \Multiply #1by #2
       {{
	\count 0 = #1\relax
	\count 2 = #2\relax
	\count 4 = 65536
	\Mess@ge {Before scaling, count 0 = \the \count 0 \space and
			count 2 = \the \count 2}%
	\ifnum	\count 0 > 32767 
	\then	\divide \count 0 by 4
		\divide \count 4 by 4
	\else	\ifnum	\count 0 < -32767
		\then	\divide \count 0 by 4
			\divide \count 4 by 4
		\else
		\fi
	\fi
	\ifnum	\count 2 > 32767 
	\then	\divide \count 2 by 4
		\divide \count 4 by 4
	\else	\ifnum	\count 2 < -32767
		\then	\divide \count 2 by 4
			\divide \count 4 by 4
		\else
		\fi
	\fi
	\multiply \count 0 by \count 2
	\divide \count 0 by \count 4
	\xdef \product {#1 = \the \count 0 \internal@nits}%
	\aftergroup \product
       }}

\def\r@duce{\ifdim\dimen0 > 90\r@dian \then   
		\multiply\dimen0 by -1
		\advance\dimen0 by 180\r@dian
		\r@duce
	    \else \ifdim\dimen0 < -90\r@dian \then  
		\advance\dimen0 by 360\r@dian
		\r@duce
		\fi
	    \fi}

\def\Sine#1%
       {{%
	\dimen 0 = #1 \r@dian
	\r@duce
	\ifdim\dimen0 = -90\r@dian \then
	   \dimen4 = -1\r@dian
	   \c@mputefalse
	\fi
	\ifdim\dimen0 = 90\r@dian \then
	   \dimen4 = 1\r@dian
	   \c@mputefalse
	\fi
	\ifdim\dimen0 = 0\r@dian \then
	   \dimen4 = 0\r@dian
	   \c@mputefalse
	\fi
	\ifc@mpute \then
		\divide\dimen0 by 180
		\dimen0=3.141592654\dimen0
		\dimen 2 = 3.1415926535897963\r@dian 
		\divide\dimen 2 by 2 
		\Mess@ge {Sin: calculating Sin of \nodimen 0}%
		\count 0 = 1 
		\dimen 2 = 1 \r@dian 
		\dimen 4 = 0 \r@dian 
		\loop
			\ifnum	\dimen 2 = 0 
			\then	\stillc@nvergingfalse 
			\else	\stillc@nvergingtrue
			\fi
			\ifstillc@nverging 
			\then	\term {\count 0} {\dimen 0} {\dimen 2}%
				\advance \count 0 by 2
				\count 2 = \count 0
				\divide \count 2 by 2
				\ifodd	\count 2 
				\then	\advance \dimen 4 by \dimen 2
				\else	\advance \dimen 4 by -\dimen 2
				\fi
		\repeat
	\fi		
			\xdef \sine {\nodimen 4}%
       }}

\def\Cosine#1{\ifx\sine\UnDefined\edef\Savesine{\relax}\else
		             \edef\Savesine{\sine}\fi
	{\dimen0=#1\r@dian\advance\dimen0 by 90\r@dian
	 \Sine{\nodimen 0}
	 \xdef\cosine{\sine}
	 \xdef\sine{\Savesine}}}	      

\def\psdraft{
	\def\@psdraft{0}
}
\def\psfull{
	\def\@psdraft{100}
}

\psfull

\newif\if@scalefirst
\def\psscalefirst{\@scalefirsttrue}
\def\psrotatefirst{\@scalefirstfalse}
\psrotatefirst

\newif\if@draftbox
\def\psnodraftbox{
	\@draftboxfalse
}
\def\psdraftbox{
	\@draftboxtrue
}
\@draftboxtrue

\newif\if@prologfile
\newif\if@postlogfile
\def\pssilent{
	\@noisyfalse
}
\def\psnoisy{
	\@noisytrue
}
\psnoisy
\newif\if@bbllx
\newif\if@bblly
\newif\if@bburx
\newif\if@bbury
\newif\if@height
\newif\if@width
\newif\if@rheight
\newif\if@rwidth
\newif\if@angle
\newif\if@clip
\newif\if@verbose
\def\@p@@sclip#1{\@cliptrue}



\def\@p@@sfigure#1{\def\@p@sfile{null}\def\@p@sbbfile{null}
	        \openin1=#1.bb
		\ifeof1\closein1
	        	\openin1=\figurepath#1.bb
			\ifeof1\closein1
			        \openin1=#1
				\ifeof1\closein1%
				       \openin1=\figurepath#1
					\ifeof1
					   \ps@typeout{Error, File #1 not found}
						\if@bbllx\if@bblly
				   		\if@bburx\if@bbury
			      				\def\@p@sfile{#1}%
			      				\def\@p@sbbfile{#1}%
				  	   	\fi\fi\fi\fi
					\else\closein1
				    		\def\@p@sfile{\figurepath#1}%
				    		\def\@p@sbbfile{\figurepath#1}%
	                       		\fi%
			 	\else\closein1%
					\def\@p@sfile{#1}
					\def\@p@sbbfile{#1}
			 	\fi
			\else
				\def\@p@sfile{\figurepath#1}
				\def\@p@sbbfile{\figurepath#1.bb}
			\fi
		\else
			\def\@p@sfile{#1}
			\def\@p@sbbfile{#1.bb}
		\fi}

\def\@p@@sfile#1{\@p@@sfigure{#1}}

\def\@p@@sbbllx#1{
		\@bbllxtrue
		\dimen100=#1
		\edef\@p@sbbllx{\number\dimen100}
}
\def\@p@@sbblly#1{
		\@bbllytrue
		\dimen100=#1
		\edef\@p@sbblly{\number\dimen100}
}
\def\@p@@sbburx#1{
		\@bburxtrue
		\dimen100=#1
		\edef\@p@sbburx{\number\dimen100}
}
\def\@p@@sbbury#1{
		\@bburytrue
		\dimen100=#1
		\edef\@p@sbbury{\number\dimen100}
}
\def\@p@@sheight#1{
		\@heighttrue
		\dimen100=#1
   		\edef\@p@sheight{\number\dimen100}
}
\def\@p@@swidth#1{
		\@widthtrue
		\dimen100=#1
		\edef\@p@swidth{\number\dimen100}
}
\def\@p@@srheight#1{
		\@rheighttrue
		\dimen100=#1
		\edef\@p@srheight{\number\dimen100}
}
\def\@p@@srwidth#1{
		\@rwidthtrue
		\dimen100=#1
		\edef\@p@srwidth{\number\dimen100}
}
\def\@p@@sangle#1{
		\@angletrue
		\edef\@p@sangle{#1} 
}
\def\@p@@ssilent#1{ 
		\@verbosefalse
}
\def\@p@@sprolog#1{\@prologfiletrue\def\@prologfileval{#1}}
\def\@p@@spostlog#1{\@postlogfiletrue\def\@postlogfileval{#1}}
\def\@cs@name#1{\csname #1\endcsname}
\def\@setparms#1=#2,{\@cs@name{@p@@s#1}{#2}}
%
%
\def\ps@init@parms{
		\@bbllxfalse \@bbllyfalse
		\@bburxfalse \@bburyfalse
		\@heightfalse \@widthfalse
		\@rheightfalse \@rwidthfalse
		\def\@p@sbbllx{}\def\@p@sbblly{}
		\def\@p@sbburx{}\def\@p@sbbury{}
		\def\@p@sheight{}\def\@p@swidth{}
		\def\@p@srheight{}\def\@p@srwidth{}
		\def\@p@sangle{0}
		\def\@p@sfile{} \def\@p@sbbfile{}
		\def\@p@scost{10}
		\def\@sc{}
		\@prologfilefalse
		\@postlogfilefalse
		\@clipfalse
		\if@noisy
			\@verbosetrue
		\else
			\@verbosefalse
		\fi
}
%
%
\def\parse@ps@parms#1{
	 	\@psdo\@psfiga:=#1\do
		   {\expandafter\@setparms\@psfiga,}}
%
%
\newif\ifno@bb
\def\bb@missing{
	\if@verbose{
		\ps@typeout{psfig: searching \@p@sbbfile \space  for bounding box}
	}\fi
	\no@bbtrue
	\epsf@getbb{\@p@sbbfile}
        \ifno@bb \else \bb@cull\epsf@llx\epsf@lly\epsf@urx\epsf@ury\fi
}	
\def\bb@cull#1#2#3#4{
	\dimen100=#1 bp\edef\@p@sbbllx{\number\dimen100}
	\dimen100=#2 bp\edef\@p@sbblly{\number\dimen100}
	\dimen100=#3 bp\edef\@p@sbburx{\number\dimen100}
	\dimen100=#4 bp\edef\@p@sbbury{\number\dimen100}
	\no@bbfalse
}
\newdimen\p@intvaluex
\newdimen\p@intvaluey
\def\rotate@#1#2{{\dimen0=#1 sp\dimen1=#2 sp
		  \global\p@intvaluex=\cosine\dimen0
		  \dimen3=\sine\dimen1
		  \global\advance\p@intvaluex by -\dimen3
		  \global\p@intvaluey=\sine\dimen0
		  \dimen3=\cosine\dimen1
		  \global\advance\p@intvaluey by \dimen3
		  }}
\def\compute@bb{
		\no@bbfalse
		\if@bbllx \else \no@bbtrue \fi
		\if@bblly \else \no@bbtrue \fi
		\if@bburx \else \no@bbtrue \fi
		\if@bbury \else \no@bbtrue \fi
		\ifno@bb \bb@missing \fi
		\ifno@bb \ps@typeout{FATAL ERROR: no bb supplied or found}
			\no-bb-error
		\fi
		%
%
		\count203=\@p@sbburx
		\count204=\@p@sbbury
		\advance\count203 by -\@p@sbbllx
		\advance\count204 by -\@p@sbblly
		\edef\ps@bbw{\number\count203}
		\edef\ps@bbh{\number\count204}
		\if@angle 
			\Sine{\@p@sangle}\Cosine{\@p@sangle}
	        	{\dimen100=\maxdimen\xdef\r@p@sbbllx{\number\dimen100}
					    \xdef\r@p@sbblly{\number\dimen100}
			                    \xdef\r@p@sbburx{-\number\dimen100}
					    \xdef\r@p@sbbury{-\number\dimen100}}
%
                        \def\minmaxtest{
			   \ifnum\number\p@intvaluex<\r@p@sbbllx
			      \xdef\r@p@sbbllx{\number\p@intvaluex}\fi
			   \ifnum\number\p@intvaluex>\r@p@sbburx
			      \xdef\r@p@sbburx{\number\p@intvaluex}\fi
			   \ifnum\number\p@intvaluey<\r@p@sbblly
			      \xdef\r@p@sbblly{\number\p@intvaluey}\fi
			   \ifnum\number\p@intvaluey>\r@p@sbbury
			      \xdef\r@p@sbbury{\number\p@intvaluey}\fi
			   }
			\rotate@{\@p@sbbllx}{\@p@sbblly}
			\minmaxtest
			\rotate@{\@p@sbbllx}{\@p@sbbury}
			\minmaxtest
			\rotate@{\@p@sbburx}{\@p@sbblly}
			\minmaxtest
			\rotate@{\@p@sbburx}{\@p@sbbury}
			\minmaxtest
			\edef\@p@sbbllx{\r@p@sbbllx}\edef\@p@sbblly{\r@p@sbblly}
			\edef\@p@sbburx{\r@p@sbburx}\edef\@p@sbbury{\r@p@sbbury}
		\fi
		\count203=\@p@sbburx
		\count204=\@p@sbbury
		\advance\count203 by -\@p@sbbllx
		\advance\count204 by -\@p@sbblly
		\edef\@bbw{\number\count203}
		\edef\@bbh{\number\count204}
}
%
%
\def\in@hundreds#1#2#3{\count240=#2 \count241=#3
		     \count100=\count240	
		     \divide\count100 by \count241
		     \count101=\count100
		     \multiply\count101 by \count241
		     \advance\count240 by -\count101
		     \multiply\count240 by 10
		     \count101=\count240	
		     \divide\count101 by \count241
		     \count102=\count101
		     \multiply\count102 by \count241
		     \advance\count240 by -\count102
		     \multiply\count240 by 10
		     \count102=\count240	
		     \divide\count102 by \count241
		     \count200=#1\count205=0
		     \count201=\count200
			\multiply\count201 by \count100
		 	\advance\count205 by \count201
		     \count201=\count200
			\divide\count201 by 10
			\multiply\count201 by \count101
			\advance\count205 by \count201
		     \count201=\count200
			\divide\count201 by 100
			\multiply\count201 by \count102
			\advance\count205 by \count201
		     \edef\@result{\number\count205}
}
\def\compute@wfromh{
		\in@hundreds{\@p@sheight}{\@bbw}{\@bbh}
		\edef\@p@swidth{\@result}
}
\def\compute@hfromw{
	        \in@hundreds{\@p@swidth}{\@bbh}{\@bbw}
		\edef\@p@sheight{\@result}
}
\def\compute@handw{
		\if@height 
			\if@width
			\else
				\compute@wfromh
			\fi
		\else 
			\if@width
				\compute@hfromw
			\else
				\edef\@p@sheight{\@bbh}
				\edef\@p@swidth{\@bbw}
			\fi
		\fi
}
\def\compute@resv{
		\if@rheight \else \edef\@p@srheight{\@p@sheight} \fi
		\if@rwidth \else \edef\@p@srwidth{\@p@swidth} \fi
}
%
\def\compute@sizes{
	\compute@bb
	\if@scalefirst\if@angle
	\if@width
	   \in@hundreds{\@p@swidth}{\@bbw}{\ps@bbw}
	   \edef\@p@swidth{\@result}
	\fi
	\if@height
	   \in@hundreds{\@p@sheight}{\@bbh}{\ps@bbh}
	   \edef\@p@sheight{\@result}
	\fi
	\fi\fi
	\compute@handw
	\compute@resv}

%
%
\def\psfig#1{\vbox {
	%
	\ps@init@parms
	\parse@ps@parms{#1}
	\compute@sizes
	\ifnum\@p@scost<\@psdraft{
		\special{ps::[begin] 	\@p@swidth \space \@p@sheight \space
				\@p@sbbllx \space \@p@sbblly \space
				\@p@sbburx \space \@p@sbbury \space
				startTexFig \space }
		\if@angle
			\special {ps:: \@p@sangle \space rotate \space} 
		\fi
		\if@clip{
			\if@verbose{
				\ps@typeout{(clip)}
			}\fi
			\special{ps:: doclip \space }
		}\fi
		\if@prologfile
		    \special{ps: plotfile \@prologfileval \space } \fi
			\if@verbose{
				\ps@typeout{psfig: including \@p@sfile \space }
			}\fi
			\special{ps: plotfile \@p@sfile \space }
		\if@postlogfile
		    \special{ps: plotfile \@postlogfileval \space } \fi
		\special{ps::[end] endTexFig \space }
		\vbox to \@p@srheight sp{
			\hbox to \@p@srwidth sp{
				\hss
			}
		\vss
		}
	}\else{
		\if@draftbox{		
			\hbox{\frame{\vbox to \@p@srheight sp{
			\vss
			\hbox to \@p@srwidth sp{ \hss \@p@sfile \hss }
			\vss
			}}}
		}\else{
			\vbox to \@p@srheight sp{
			\vss
			\hbox to \@p@srwidth sp{\hss}
			\vss
			}
		}\fi

	}\fi
}}
\psfigRestoreAt
\let\@=\LaTeXAtSign
\def\FigureInsert#1%
{{%
    \parindent = 0pt \leftskip = 0pt \rightskip = 0pt%
    \vskip .4in%
    \leavevmode%
    \centerline{\psfig{figure=#1,clip=t}}%
    \nobreak%
    \vglue .1in%
    \nobreak%
}}
\def\FigureInsert#1#2%
{{%
    \def\CompareStrings##1##2%
    {%
        TT\fi%
        \edef\StringOne{##1}%
        \edef\StringTwo{##2}%
        \ifx\StringOne\StringTwo%
    }%
    \parindent = 0pt \leftskip = 0pt \rightskip = 0pt%
    \vskip .4in%
    \leavevmode%
    \if\CompareStrings{#2}{left}%
        \leftline{\psfig{figure=#1,clip=t}}%
    \else\if\CompareStrings{#2}{center}%
        \centerline{\psfig{figure=#1,clip=t}}%
    \else\if\CompareStrings{#2}{right}%
        \rightline{\psfig{figure=#1,clip=t}}%
    \fi\fi\fi%
    \nobreak%
    \vglue .1in%
    \nobreak%
}}
%

\global\def\FigureInsertScaled#1#2#3%
{{%
    \def\CompareStrings##1##2%
    {%
        TT\fi%
        \edef\StringOne{##1}%
        \edef\StringTwo{##2}%
        \ifx\StringOne\StringTwo%
    }%
    \parindent = 0pt \leftskip = 0pt \rightskip = 0pt%
    \vskip .4in%
    \leavevmode%
    \if\CompareStrings{#2}{left}%
        \leftline{\psfig{figure=#1,height=#3,clip=t}}%
    \else\if\CompareStrings{#2}{center}%
        \centerline{\psfig{figure=#1,height=#3,clip=t}}%
    \else\if\CompareStrings{#2}{right}%
        \rightline{\psfig{figure=#1,height=#3,clip=t}}%
    \fi\fi\fi%
    \nobreak%
    \vglue .1in%
    \nobreak%
}}

%

%

%
%
\else
    \def\Flush{}
    \def\beginmini{\FillMode\BeginMiniGeneral}
    \def\minihead#1{} 
    \def\miniheadinsert#1{} 
\fi
{\tenpoint\mib\relax}
{\twelvepoint\mib\relax\rm}
\extraminifonts
\def\BottomCaptionInsert#1{%
    \topinsert%
    \BottomCaption{#1}%
    \endinsert%
    \vfill\eject\vglue 9in\vfill\eject%
}
\def\WideCaptionInsert#1{%
    \topinsert%
    \WideCaption{#1}%
    \endinsert%
    \vfill\eject\vglue 9in\vfill\eject%
}
\def\SideCaptionInsert#1{%
    \topinsert%
    \SideCaption{#1}%
    \endinsert%
    \vfill\eject\vglue 9in\vfill\eject%
}
\long\def\BottomCaption#1{%
    \setbox1=\hbox to 9.5in{\hss\minifigbox{}{9in}{9in}\hss}%
    \twelvepoint\baselineskip 14pt%
    \setbox2=\hbox to 9.5in{\hss\vtop{\hsize = 9in\noindent #1\par}\hss}%
    \wd1=0pt%
    \wd2=0pt%
    \box1%
    \vglue .8in%
    \box2%
}
\long\def\WideCaption#1#2#3#4{%
    \setbox1=\hbox to 9.5in{\hss\minifigbox{}{#1}{#2}\hss}%
    \twelvepoint\baselineskip 14pt%
    \setbox2=\hbox to 9.5in{\hss\vtop{\hsize = #3\noindent #4\par}\hss}%
    \wd1=0pt%
    \wd2=0pt%
    \box1%
    \vglue .1in%
    \box2%
}
\long\def\SideCaption#1{%
    \setbox1=\hbox to 9.5in{\hss\minifigbox{}{9in}{9in}\hss}%
    \twelvepoint\baselineskip 14pt%
    \setbox2=\hbox{\vtop{\hsize = 3.8in\noindent #1\par}}%
    \wd1=0pt%
    \wd2=0pt%
    \setbox3=\hbox to 9.5in{\hglue 5in\box2\hss}%
    \wd3=0pt%
    \box1%
    \vglue -4in%
    \box3%
}
%
\newdimen\strutskip
\newskip\ttglue
\def\strut{\vrule height 0.8\strutskip depth 0.3\strutskip width 0pt}
\def\phh{\phantom{11}\relax}
\def\pmalign#1#2{$\llap{$#1$}\pm\rlap{$#2$}$}
\def\Ast{\hbox{$\ast$}}
\def\twoast{\hbox{\Ast\Ast}}
\def\fourast{\hbox{\Ast\Ast\Ast\Ast}}
\def\threeast{\hbox{\Ast\Ast\Ast}}
\def\pma#1#2{$\llap{$#1$}\pm\rlap{$#2$}$}
\def\pmb#1#2{$\llap{$#1$}\phantom{\>\pm\>}\rlap{$#2$}$}
\def\ealign#1#2{$\llap{$#1$}=\rlap{$#2$}$}
%
\newdimen\miniLinewidth                \miniLinewidth=0.001in
\def\miniBoxit#1#2#3%
{%
    \vbox%
    {%
        \hrule height \miniLinewidth%
        \hbox%
        {%
            \vrule width \miniLinewidth%
            \vtop%
            {%
                \kern #3
                \hbox%
                {%
                    \kern #2
                    \vbox{\hbox to 0in{\hss\copy#1\hss}}%
                    \kern #2
                }%
                \kern #3
            }%
            \vrule width \miniLinewidth%
        }%
        \hrule height \miniLinewidth%
    }%
}
\newdimen\minifigboxwidth                \minifigboxwidth=4.25in
\newdimen\minifigboxheight                \minifigboxheight=4.25in
\def\minifigbox#1#2#3{%
    \setbox0=\hbox{#1}%
    \dp0=0pt%
    \ht0=0pt%
    \minifigboxwidth=#2\relax%
    \minifigboxheight=#3\relax%
    \divide\minifigboxwidth by 2\relax%
    \divide\minifigboxheight by 2\relax%
    \miniBoxit{0}{\minifigboxwidth}{\minifigboxheight}%
}
\def \bye{\vfil\eject\end}
\def \sih{a-Si:H}
\def \q{\hskip 0.25in}
\def \ph#1{\phantom{#1}}
\def \\{$\backslash$}
\def\ftstrut{\vrule height 5pt depth 0pt width 0pt}
\def\frac#1#2{{\displaystyle{#1 \over #2}}}
\def\textfrac#1#2{{\textstyle{#1 \over #2^{\ftstrut}}}}
\def\SingleRule{\hrule width \hsize}
\def\SingleRuleMid{\nobreak\smallskip\SingleRule\nobreak\smallskip}
\def\DoubleRule{\SingleRule\vglue2pt\SingleRule}
\def\DoubleRuleMid{\nobreak\medskip\DoubleRule\nobreak\medskip}
\def\DoubleRuleEnd{\nobreak\medskip\DoubleRule}
\hyphenation{%
    brems-strah-lung
    Dan-ko-wych
    Fuku-gi-ta
    Gav-il-let
    Gla-show
    mono-pole
    mono-poles
    Sad-ler
}
%
%
%
\let\miniuppercase=\uppercase
\let\mininobreak=\nobreak
\let\mininoindent=\noindent
\let\minimedskip=\medskip
\let\minihfill=\hfill
\let\minibaselineskip=\baselineskip
\ifnum\MiniFinal = 0
    \def\IndexPageno#1{ }
    \def\IndexText#1#2{ }
    \def\nominititle#1{%
        \twelvepoint%
        \noindent Orange Index Term = #1\par\bigskip%
        \noindent THERE IS NO TITLE\par\bigskip%
        \mininobreak\twelvepoint\minibaselineskip=17pt\relax }
    \def\minititle#1{%
        \twelvepoint%
        \noindent Orange Index Term = #1\par\bigskip%
        \twelvepoint\bf\mib%
        \bgroup\mininoindent\miniuppercase{#1}%
        \egroup\minihfill\mininobreak\minimedskip%
        \mininobreak\twelvepoint\minibaselineskip=17pt}
\else
    \def\nominititle#1{%
        \twelvepoint\minibaselineskip=17pt\relax }
    \def\minititle#1{%
        \twelvepoint\bf\mib%
        \bgroup\mininoindent\miniuppercase{#1}\egroup%
        \minihfill\mininobreak\minimedskip%
        \mininobreak\twelvepoint\minibaselineskip=17pt}
\fi
\ifnum\Miniwww=3 
\advance\hoffset by 1in
\def\nominititle#1{%
        \twelvepoint%
        \mininobreak\twelvepoint\minibaselineskip=17pt\relax }
    \def\minititle#1{%
        \twelvepoint%
        \twelvepoint\bf\mib%
        \bgroup\mininoindent\miniuppercase{#1}%
        \egroup\minihfill\mininobreak\minimedskip%
        \mininobreak\twelvepoint\minibaselineskip=17pt}
\newbox\wwwfootcitation
\gdef\runningdate{\bgroup\sevenrm\today\quad\TimeOfDay\egroup}
\def\minihead#1%
    {%
     \global\headline={\bf \hss-- \the\pageno --\hss}%
     \global\footline={\bf \hss-- \the\pageno --\hss}%
     \global\footline={\bgroup\sevenrm \hfill\today\quad\TimeOfDay\hfill\egroup}%
    \setbox\wwwfootcitation=\vtop{%
   	\vglue -.1in%
   	\hbox to  6.0in{%
	      \hss{\sevenrm CITATION: S.~Eidelman~{\sevenit et al.},
	      Phys.~Lett.~B~{\sevenbf 592}, 1 (2004)
       (URL: http://pdg.lbl.gov/)}\hss}
	 \vglue .005in%
	 \hbox to  6in{%
	      \hss{
      	      \qquad\runningdate}\hss}
   \vss%
             }%
\footline={\ifnum\pageno=1\firstfoot\else\restoffoot\fi}
\gdef\firstfoot{\centerline{\hss\copy\wwwfootcitation\hss}}
\gdef\restoffoot{\centerline{\hss\runningdate\hss}}
    }
\fi
\def\miniauthor#1{%
        \nobreak\twelvepoint\noindent #1%
        \hfill\nobreak\medskip\nobreak\twelvepoint\baselineskip=17pt}
\def\breakminititle{\hfill\break\nobreak\noindent }
\def\breakminiauthor{\hfill\break\nobreak\noindent }
\def\miniref{\bigskip\parindent=20pt\leftline{\bf Reference}%
        \baselineskip=14pt\nobreak\smallskip\nobreak}
\def\minirefs{\bigskip\parindent=20pt\leftline{\bf References}%
        \baselineskip=14pt\nobreak\smallskip\nobreak}
\def\mininotesrefs{\bigskip\parindent=20pt\leftline{\bf Notes and References}%
        \baselineskip=14pt\nobreak\smallskip\nobreak}
\def\minifootrefs{\bigskip\parindent=20pt%
        \leftline{\bf Footnotes and References}%
        \baselineskip=14pt\nobreak\smallskip\nobreak}
\def\minirefsI{\bigskip\parindent=20pt\leftline{\bf References for Section~I}%
        \baselineskip=14pt\nobreak\smallskip\nobreak}
\def\minirefsIandII{\bigskip\parindent=20pt\leftline{\bf References for Sections~I~and~II}%
        \baselineskip=14pt\nobreak\smallskip\nobreak}
\def\minirefsII{\bigskip\parindent=20pt\leftline{\bf References for Section~II}%
        \baselineskip=14pt\nobreak\smallskip\nobreak}
\def\minirefsIII{\bigskip\parindent=20pt\leftline{\bf References for Section~III}%
        \baselineskip=14pt\nobreak\smallskip\nobreak}
\def\minirefsIV{\bigskip\parindent=20pt\leftline{\bf References for Section~IV}%
        \baselineskip=14pt\nobreak\smallskip\nobreak}
\def\minirefsV{\bigskip\parindent=20pt\leftline{\bf References for Section~V}%
        \baselineskip=14pt\nobreak\smallskip\nobreak}